\documentclass[aip,amsmath,amssymb,reprint]{revtex4-1}

\usepackage{graphicx}
\usepackage{dcolumn}
\usepackage{bm}

\usepackage[utf8]{inputenc}
\usepackage[T1]{fontenc}
\usepackage{mathptmx}
\usepackage{etoolbox}

\usepackage{float}
\usepackage{stmaryrd}
\usepackage{tipa}
\usepackage{bbold}
\usepackage{graphicx}
\usepackage[svgnames]{xcolor}
\usepackage{array}
\usepackage{multirow}
\usepackage{mathtools}
\usepackage[colorlinks=true,
            linkcolor=cyan,
            citecolor=cyan,
            urlcolor=cyan]{hyperref}
\usepackage{ulem}
\usepackage{makecell}
\setcellgapes{10pt}

\usepackage{booktabs}
\usepackage{colortbl}

\usepackage{physics}

\usepackage{textgreek}

\definecolor{gg}{RGB}{8, 135, 68}

\newcommand{\mathbbm}[1]{\text{\usefont{U}{bbm}{m}{n}#1}}
\newcommand{\Id}{\mathbbm{1}} 

\usepackage{comment}

\makeatletter
\def\@email#1#2{%
 \endgroup
 \patchcmd{\titleblock@produce}
  {\frontmatter@RRAPformat}
  {\frontmatter@RRAPformat{\produce@RRAP{*#1\href{mailto:#2}{#2}}}\frontmatter@RRAPformat}
  {}{}
}%
\makeatother

\begin{document}


\title{Entanglement transfer during quantum frequency conversion in gas-filled hollow-core fibers} 




\author{Tasio Gonzalez-Raya}
\email{tgonzalez@bcamath.org}
\thanks{The first two authors contributed equally to this work.}
\affiliation{Basque Center for Applied Mathematics (BCAM), Alameda de Mazarredo 14, 48009 Bilbao, Spain}
\affiliation{EHU Quantum Center, University of the Basque Country UPV/EHU, Bilbao, Spain}
\author{Arturo Mena}
\thanks{The first two authors contributed equally to this work.}
\affiliation{Department of Communications Engineering, Engineering School of Bilbao, University of the Basque Country UPV/EHU, Torres Quevedo 1, 48013 Bilbao, Spain}
\author{Miriam Lazo}
\affiliation{Department of Physical Chemistry, University of the Basque Country UPV/EHU, Apartado 644, 48080 Bilbao, Spain}
\author{Luca Leggio}
\affiliation{Basque Center for Applied Mathematics (BCAM), Alameda de Mazarredo 14, 48009 Bilbao, Spain}
\affiliation{Department of Communications Engineering, Engineering School of Bilbao, University of the Basque Country UPV/EHU, Torres Quevedo 1, 48013 Bilbao, Spain}
\author{David Novoa}
\affiliation{EHU Quantum Center, University of the Basque Country UPV/EHU, Bilbao, Spain}
\affiliation{Department of Communications Engineering, Engineering School of Bilbao, University of the Basque Country UPV/EHU, Torres Quevedo 1, 48013 Bilbao, Spain}
\affiliation{IKERBASQUE, Basque Foundation for Science, Plaza Euskadi 5, 48009 Bilbao, Spain}
\author{Mikel Sanz}
\affiliation{Basque Center for Applied Mathematics (BCAM), Alameda de Mazarredo 14, 48009 Bilbao, Spain}
\affiliation{EHU Quantum Center, University of the Basque Country UPV/EHU, Bilbao, Spain}
\affiliation{Department of Physical Chemistry, University of the Basque Country UPV/EHU, Apartado 644, 48080 Bilbao, Spain}
\affiliation{IKERBASQUE, Basque Foundation for Science, Plaza Euskadi 5, 48009 Bilbao, Spain}




\begin{abstract}
Quantum transduction is essential for the future hybrid quantum networks, connecting devices across different spectral ranges. In this regard, molecular modulation in hollow-core fibers has proven to be exceptional for efficient and tunable frequency conversion of arbitrary light fields down to the single-photon limit. However, insights on this conversion method for quantum light have remained elusive beyond standard semiclassical models. In this Letter, we employ a quantum Hamiltonian framework to characterize the behavior of entanglement during molecular modulation, while describing the quantum dynamics of both molecules and photons in agreement with recent experiments. In particular, apart from obtaining analytical expressions for the final opto-molecular states, our model predicts a close correlation between the evolution of the average photon numbers and the transfer of entanglement between the interacting parties. These results will contribute to the development of new fiber-based strategies to tackle the challenges associated with the upcoming generation of lightwave quantum technologies. 

\end{abstract}


\maketitle 



%
%

%

\section{Introduction}

Understanding light-matter interactions at the quantum level lies at the core of the recent developments in quantum technologies~\cite{Tavis1968, Aspelmeyer2014, Schmidt2016, Blais2021} that are behind sophisticated systems such as the future hybrid quantum networks~\cite{Kimble2008}. These systems comprise multiple devices such as quantum light sources\cite{Arakawa2020,Zhang2024}, repeaters\cite{Munro2015,Azuma2023}, memories~\cite{Heshami2016,Lei2023}, fiber transmission lines~\cite{Wang2022}, etc., which operate across different spectral regions of the optical domain, in sharp contrast to e.g. the microwave superconducting circuits employed in state-of-art quantum computers~\cite{Girvin2014,Gu2017}. Thus, efficient frequency transduction of quantum light states between disparate domains ~\cite{Li2004,Radnaev2010,Lauk2020} is essential to bridge the operational gaps between nodes~\cite{Awschalom2021}. 
This has encouraged the proposal and demonstration of many different strategies to tackle this challenge in different platforms~\cite{Gnauck2006,Gogyan2008,McGuinness2011,Ramelow2012,Clark2013,Williamson2014,Donvalkar2014,Lefrancois2015,Bell2016,Fan2016,Clemmen2018,Bonsma2022}. Recently, molecular modulation in hollow-core anti-resonant fibers (ARFs) filled with gas~\cite{Russell2014,Numkam2023} has stood out owing to its tunability, large frequency shifts, near-unity efficiency and exquisite preservation of non-classical correlations~\cite{Sokolov2022, Tyumenev2022,Hamer2024fiber}. This is facilitated by the tight light-matter confinement in the core~\cite{Russell2014}, ultralow attenuation over a broad bandwidth~\cite{Numkam2023} and pressure-adjustable optical properties~\cite{Bauerschmidt2015}, which make ARFs excellent vehicles for light-based quantum applications~\cite{Finger2015,Antesberger2024}.

On the other hand, in molecular modulation~\cite{Liang2000,Weber2012} at the single-photon level, a quantum light state scatters off the molecular coherence waves pre-excited via stimulated Raman scattering (SRS) in the ARF core, changing its frequency by the appropriate Raman shift without threshold. The corresponding state can be controllably up- or down-converted provided specific phase-matching conditions are fulfilled, which in the case of gas-filled ARFs is achieved by leveraging the fiber dispersion~\cite{Hosseini2017, Mridha2019}.

Despite the great potential of ARF-based molecular modulation for quantum transduction~\cite{Tyumenev2022}, it still remains unclear whether intrinsic quantum properties such as entanglement can be transferred with high fidelity from the original to the target states during the conversion process, a question that cannot be answered using the widely-employed Maxwell-Bloch formalism~\cite{Raymer1985, Raymer1990}. The main reason is its classical treatment of the light fields, although it has been applied to the modelling of certain quantum optical phenomena like photon absorption and emission in weakly-excited atomic clouds~\cite{Svidzinsky2015}. Recent efforts in this direction have provided further insights on the changes in photonic correlations after frequency conversion~\cite{Wang2023}. Nevertheless, a more detailed description of the internal quantum light-quantum matter interactions down to single-photon limit has, to our best knowledge, so far not been adapted to molecular modulation-based frequency conversion in ARFs.




In this Letter, we describe both the preparation of the molecular state through SRS, as well as the subsequent thresholdless frequency-conversion process at the single-photon level by employing a quantum Hamiltonian model that allows us to study the behavior of entanglement during molecular modulation. In particular, considering the experimental scenario reported in Ref.~\onlinecite{Tyumenev2022}, we are able to characterize the state of the molecules and predict a complete transfer of entanglement between one of the frequencies of a Bell state and its corresponding frequency-converted counterpart. Our results may aid the design, optimization and interpretation of future experiments in light-based quantum technologies using ARFs and their subsequent applications.


\section{Quantum description of molecular modulation}

\subsection{Preparation of the molecular state}

\begin{figure}[h]
{\includegraphics[width=\linewidth]{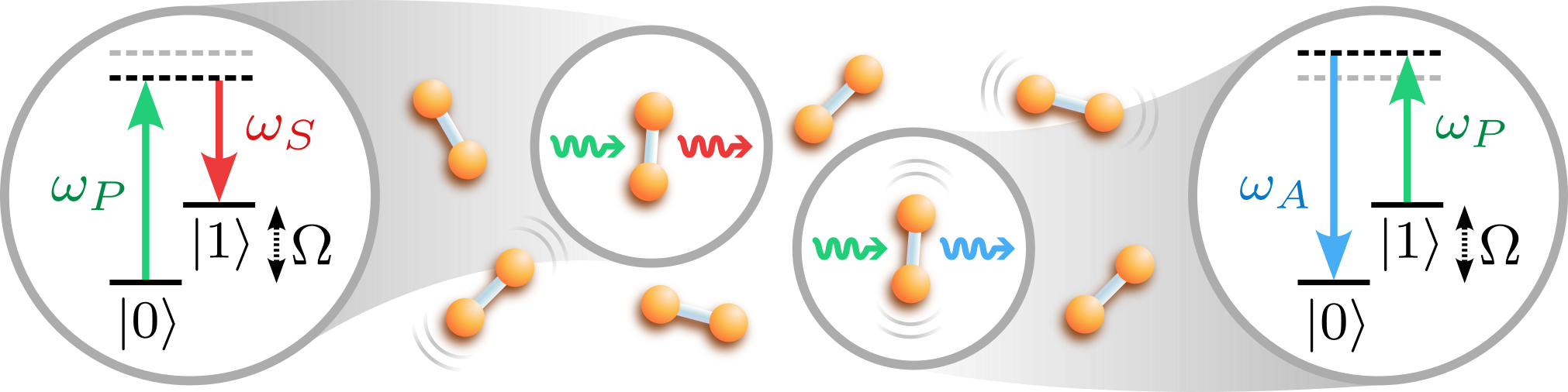}}
\caption{Illustration of Raman scattering in a molecular diatomic gas. (Left) A photon with pump frequency $\omega_P$ is inelastically scattered by a molecule in the vibrational ground state $|0\rangle$. As a result of the interaction, the molecule gains an energy defined by the Raman frequency $\Omega$, transitioning into the excited vibrational state $|1\rangle$. Meanwhile, the scattered photon ends with Stokes frequency $\omega_S=\omega_P-\Omega$. (Right) The inverse process is also represented, involving the de-excitation of molecules via the inelastic scattering of a pump photon into the anti-Stokes frequency $\omega_{A}=\omega_P +\Omega$. Dashed lines indicate off-resonant energy levels.}
\label{fig1}
\end{figure}

\begin{figure}[h]
{\includegraphics[width=\linewidth]{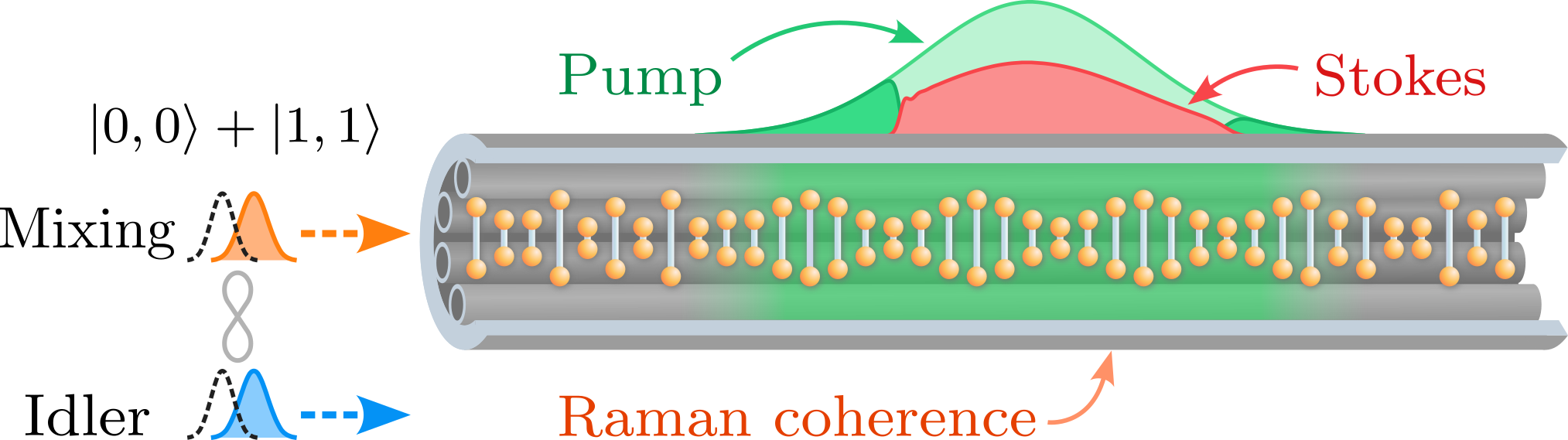}}
\caption{Schematic representation of the experimental layout considered. A pump beam generates Raman vibrational excitations in the gas molecules, preparing them in a coherent and synchronized vibrational motion. During this process, the pump is depleted into the Stokes frequency, as represented in the figure. The mixing signal simultaneously propagating with the pump perceives the molecular coherence wave and it is scattered to a higher frequency. Although the gas molecules are depicted equidistantly along a longitudinal axis for illustrative purposes, they actually fill the whole interior of the fiber and are randomly located and oriented.}
\label{fig2}
\end{figure}

The process we want to describe is molecular modulation assisted by SRS~\cite{Harris1998,Nazarkin1999}. We consider two-level molecules in general and, inspired by the recent experiments~\cite{Tyumenev2022,Aghababaei2023,Hamer2024,Hamer2024fiber}, we focus on the $Q(1)$ vibrational transition of hydrogen as a good two-level approximation. This transition is dipole-forbidden, and therefore needs a two-photon process such as Raman scattering to occur (depicted in Fig.~\ref{fig1}). 
In this process, the pump photons launched in the fundamental core mode of the H$_2$-filled ARF (illustrated in Fig.~\ref{fig2}) are scattered into the Stokes or anti-Stokes frequencies depending on whether they excite or de-excite the molecules, respectively. These transitions are illustrated in Fig.~\ref{fig1}, where $\omega_{P}$, $\omega_{S}$, and $\omega_{A}$ are the central angular frequencies of the narrowband pump, Stokes, and anti-Stokes signals, and $\Omega$ represents the Raman shift, such that $\omega_{S}=\omega_{P}-\Omega$ and $\omega_{A}=\omega_{P}+\Omega$. For the H$_2$ gas case, $\Omega/2\pi \approx 125 \text{ THz}$~\cite{Veirs1987}, the largest molecular shift in nature. Without loss of generality, the light is linearly polarized in our analysis, and therefore rotational states are highly disfavored. Furthermore, hereafter we will consider all the optical frequencies involved in the dynamics contained in the fundamental transmission band of the ARF, i.e. spectrally away from loss-inducing resonances with spatial modes localized in the cladding elements~\cite{Numkam2023}.

The quantum Hamiltonian describing the pump, Stokes, and anti-Stokes frequencies of the electric field interacting with a single molecule is expressed as $H=H_{0}+V$~\cite{Scully1997,Begzjav2022}. On the one hand, we have the unperturbed part of the Hamiltonian $H_0$, defined as
\begin{equation}
H_{0}/\hbar = \omega_{0} |0\rangle\langle0| + \omega_{1} |1\rangle\langle1| + \sum_{i=2}^{\infty}\omega_{i} |i\rangle\langle i| + \sum_{l}\omega_{l}a^{\dagger}_{l}a_{l},
\end{equation}
 where $l\in\{\text{P, S, A}\}$ labels the operators associated to the pump, Stokes, and anti-Stokes frequencies, respectively, and $\omega_{l}$ are the photonic frequencies. The quantity $\hbar\omega_{i}$ is the energy associated to the vibrational states $|i\rangle$ of the molecule. On the other hand, the interaction part of the Hamiltonian, $V$, is given by
\begin{equation}
V = \sum_{i,j}\sum_{l} g_{i,j}^{l} |i\rangle\langle j| \left(a_{l}e^{i\beta_{l}z} - a_{l}^{\dagger}e^{-i \beta_{l}z}\right),
\end{equation}
where $g_{i,j}^{l}$ is the coupling strength between levels $|i\rangle$ and $|j\rangle$ via the bosonic mode $l$, and $\beta_{l}$ is the propagation constant for frequency $\omega_l$. Let us now eliminate the higher energy levels, i.e. levels with $i>1$, to obtain an effective Hamiltonian describing the interaction of the pump, Stokes, anti-Stokes frequencies, with the molecule. In order to do this, we go to an interaction picture with respect to $H_{0}$ and perform a rotating-wave approximation, keeping only the static terms up to second order in the coupling strength. That is, by assuming that $g_{i,j}^{l}\sqrt{N_{l}}\ll \hbar |\omega_{i}-\omega_{j} \pm \omega_{l}|$, with $N_{l}$ the number of photons with frequency $\omega_l$, we keep only the resonant terms. The relevant resonances in this system are $\omega_{P}-\omega_{S} = \Omega =  \omega_{A}-\omega_{P}$, with $\Omega\equiv\omega_{1}-\omega_{0}$. Extending this approach to a system with $N$ molecules results in the following effective Hamiltonian~\cite{Walls1971}:
\begin{eqnarray}\label{H_eff}
\nonumber H_{\text{eff}} &=& \hbar\Omega J_{z} + \hbar\sum_{l} \left( \omega_{l}+\Delta_{l}^{+} \right)a^{\dagger}_{l}a_{l} + 2\hbar\sum_{l} \Delta_{l}^{-}a^{\dagger}_{l}a_{l} J_{z} \\
\nonumber &+& \hbar\left( G_{S}e^{i\Delta\beta  z} a_{P}a^{\dagger}_{S} + G_{A} e^{i\Delta\beta' z}a^{\dagger}_{P}a_{A}\right)J^{+} \\
&+& \hbar\left( G_{S}^* e^{-i\Delta\beta  z}a^{\dagger}_{P}a_{S} + G_{A}^* e^{-i\Delta\beta' z} a_{P}a^{\dagger}_{A} \right)J^{-},
\end{eqnarray}
where $G_{S(A)}$ is the interaction strength between pump, (anti-) Stokes, and the molecules, $\Delta_{l}^{\pm}$ represent the Stark shifts, and $\Delta\beta \equiv \beta_{P}-\beta_{S}$ and $\Delta\beta'\equiv \beta_{A}-\beta_{P}$. The global spin operators are defined through the $1/2$-spin operators as $J_{z} = \oplus_{l=1}^N \sigma^{z}_{l}/2$ and $J^{\pm} = \oplus_{l=1}^N \sigma^{\pm}_l$, with $\sigma^{z}=|1\rangle\langle1| - |0\rangle\langle0|$, $\sigma^{+}=|1\rangle\langle0|$, and $\sigma^{-}=|0\rangle\langle1|$. They satisfy $[J_{z},J^{\pm}]=\pm J^{\pm}$ and $[J^{+},J^{-}]=2J_{z}$ and, as operators, they act on global spin states of the form $\left|{\bf N/2},m_{z}\right\rangle$, with $m_{z}\in\{-N/2,\ldots,N/2\}$. Note that these operators treat the molecular gas as an ensemble of two-level systems and they are {\it not} representing actual angular momentum of the molecules or light polarization. More details on the derivation of $H_{\text{eff}}$ can be found in the supplementary material.

The excitation of molecular coherence manifests itself as a synchronous oscillation of the gaseous core. This is shown in Fig.~\ref{fig2}, where the molecules are depicted equidistantly along a longitudinal axis with the same orientation for illustrative purposes, whereas in reality the gas fills the whole interior of the fiber and the molecules are randomly oriented. In this regard, the interaction will be more significant with the percentage of molecules aligned with the linearly polarized fields, which is already taken into account in the coupling constants obtained phenomenologically (see supplementary material). In Ref.~\onlinecite{Tyumenev2022}, the molecular coherence was generated by a nanosecond pump pulse with an energy of 115 \textmu J. This means that the initial state of the pump can be considered as a coherent state with $\alpha_{P} \approx 2.48\times10^{7}$, enabling a few approximations that simplify Eq.~\eqref{H_eff}. 
Firstly, the annihilation of pump photons leads to the generation of intense laser radiation at the down-shifted Stokes frequency along the fiber length~\cite{Benabid2002}. Therefore, we may consider a semiclassical approximation, replacing the operators in both pump and Stokes frequencies by classical variables $a_{P}\rightarrow\alpha_{P}$ and $a_{S}\rightarrow\alpha_{S}$ in Eq.~\eqref{H_eff}. Additionally, since the majority of molecules remain in their ground state, the photon population of the anti-Stokes frequency is usually negligible in this process, allowing us to discard it in our analysis.
Finally, we consider that the Stark shifts are also negligible, $\Delta_{P}^{\pm}=\Delta_{S}^{\pm}\approx0$. Before performing the semiclassical approximation, in order to avoid oscillations with $\Omega$ in the expectation values, we transform the Hamiltonian into an interaction picture with respect to $\Omega J_{z} + \omega_{P}a_{P}^{\dagger}a_{P} + \omega_{S}a_{S}^{\dagger}a_{S}$, obtaining
\begin{equation}\label{H_alpha_I}
H_{\alpha}^{I} = \hbar\left( G_{S}e^{i\Delta\beta z}\alpha_{P}\alpha_{S}^* J^{+} + G_{S}^{*}e^{-i\Delta\beta z}\alpha_{P}^* \alpha_{S} J^{-}\right).
\end{equation}
By evolving the initial state of the molecules $\left | {\bf \frac{N}{2}},-\frac{N}{2}\right\rangle$, which corresponds to all molecules in the ground state, under this Hamiltonian for a time $t$, we find the state in the interaction picture
\begin{equation}\label{spin_coherent_state}
|s\rangle = \left(\frac{e^{-i\Omega t}}{1+|s|^{2}}\right)^{\frac{N}{2}}\sum_{n=0}^{N}\begin{pmatrix} N \\ n\end{pmatrix}^{1/2} s^{n}\left | {\bf \frac{N}{2}},-\frac{N}{2}+n\right\rangle.
\end{equation}
Here, we have
\begin{equation}
s = -ie^{i\Delta\beta  z}\tan\left(G_{S}\alpha_{P}\alpha_{S} t\right),
\end{equation}
where we have assumed that $G_{S}$, $\alpha_{P}$, and $\alpha_{S}$ are real. The supplementary material and Ref.~\onlinecite{Arecchi1972} include the mathematical steps to find $|s\rangle$. 



The emergence of vibrational coherence in the molecular gas, highlighted in green inside the fiber at Fig.~\ref{fig2}, originates from the beating between the pump and Stokes fields~\cite{Mridha2019, Raymer1990, Chen2024}. As the amplitude of the coherence wave rises, the pump starts to suffer depletion and the Stokes starts to be amplified. As the depletion continues, the beating between the fields becomes weaker, preventing the generation of new coherence. Meanwhile, the existing coherence wave fades away due to collisional dephasing on a time scale $T_2$. However, in a temporal reference frame moving with the co-propagating pump and Stokes pulses at their group velocity, new coherence will be found over a longer time scale as light travels through the fiber, since the amplitude of the coherence wave at a given relative time coordinate is the result of the integrated effects induced by these pulses at previous times~\cite{Raymer1990}. Hence, the excited molecular coherence is harvested, within its lifetime, for frequency conversion of an arbitrary mixing signal. Unlike usual SRS, in which the power of the scattered pump beam needs to be higher than a minimum threshold, this frequency conversion process is thresholdless and hence, it can be applied to a single photon. Additionally, the phase-matching conditions governing the feasibility of frequency conversion of the mixing signal are highly influenced by the dispersion contributions from both the gas and the geometry of the waveguide~\cite{Hosseini2017,Mridha2019}. In a nutshell, if the difference in the propagation constants of the ARF modes at the original and converted photon frequencies matches the propagation constant of the molecular coherence wave (given by the difference between those of the pump and Stokes fields), energy will be efficiently exchanged during the scattering event, resulting in a modification of the photon frequency according to the molecular Raman shift. 

\subsection{Mixing frequency conversion}

In the following, let us use the developed framework to analyze the frequency conversion process of one of the frequency components of a maximally-entangled Bell state. Indeed, we will convert the mixing frequency by launching it simultaneously with the pump beam, while the idler frequency remains unperturbed outside of the fiber, and observe the entanglement dynamics between the idler and the mixing and up-converted frequencies. Considering the conditions of Ref.~\onlinecite{Tyumenev2022}, the experimental system undergoes a phase-matched transition between a mixing frequency of $\approx 210$ THz (1425 nm in wavelength) and an up-converted frequency of $\approx 335$ THz (895 nm). Even though this is not strictly a Raman transition, it can be described using a Hamiltonian with the same structure as in Eq.~\eqref{H_eff}. This time, the terms describing the mixing to up-converted interaction mimic those of the pump to anti-Stokes, but with the appropriate parameters.

While the idler and the mixing fields are initially prepared in a Bell state, the up-converted field is in the vacuum state; therefore, the initial state is simply $(|0,0,0\rangle+|1,1,0\rangle)/\sqrt{2}$. This notation represents, from left to right and separated by commas, the number of photons in the idler, mixing, and up-converted frequencies, respectively. Since in the experiments\cite{Tyumenev2022, Bauerschmidt2015, Mridha2019} we typically have a large number of molecules ($\sim10^{18}$), the energy state of the molecular ensemble will not change significantly due to the introduction of a single excitation into the system. Thus, in the Hamiltonian characterizing the frequency-conversion process, we replace the global spin operators by their expectation values over the spin coherent state in Eq.~\eqref{spin_coherent_state}. That is, we replace $J^{+}\rightarrow\xi^{*}$ and $J^{-}\rightarrow\xi$, with 
\begin{equation}
\xi = \langle s|J^{-}|s\rangle = -i e^{i\Delta\beta z } \frac{N}{2} \sin\left( 2G_{S}\alpha_{P}\alpha_{S}t\right).
\end{equation}
Before continuing, let us transform the Hamiltonian into an interaction picture again with respect to $\Omega J_{z} + \omega_{M}a_{M}^{\dagger}a_{M} + \omega_{U}a_{U}^{\dagger}a_{U}$, where $\omega_{M}$ and $\omega_{U}$ are the mixing and up-converted angular frequencies. Then, the resulting Hamiltonian is
\begin{eqnarray}\label{H_xi_I}
H_{\xi}^{I} &=& \hbar G_{U}\Big( \xi^* e^{i[(\Omega + \omega_{M}-\omega_{U}) t - (\beta_{M}-\beta_{U})z]}a_{M}^{\dagger} a_{U} \\ 
\nonumber &+& \xi e^{-i[(\Omega + \omega_{M}-\omega_{U}) t - (\beta_{M}-\beta_{U})z]}a_{M} a_{U}^{\dagger}\Big),
\end{eqnarray}
where $G_{U}$ represents the interaction strength between the mixing, the up-converted, and the molecules, and is supposed to be real. If we assume that, as in the experiments, the phase matching and resonance conditions are satisfied, we would have $\beta_{U}-\beta_{M}=\Delta\beta\equiv\beta_{P}-\beta_{S}$ and $\omega_{U}-\omega_{M}=\Omega\equiv\omega_{P}-\omega_{S}$, respectively. Note that we do not necessarily require $\omega_l=\omega_m$ or $\beta_l=\beta_m$ with $l\in\{P,S\}$ and $m\in\{M,U\}$ and that, experimentally, these conditions can be adjusted to carry out frequency down-conversion as well, owing to the excellent tunability of gas-filled ARFs~\cite{Mridha2019, Hosseini2017}. 
Note that this Hamiltonian is similar to others found in the context of quantum frequency conversion~\cite{Tucker1969,Kumar1990}. Using these conditions and $H_{\xi}^{I}$ to evolve the Bell state considered for the idler and the mixing photons we obtain a state that, when transformed back to the Schr\"{o}dinger picture, becomes
\begin{multline}
     |\psi \rangle = \frac{1}{\sqrt{2}} \Big(|0,0,0\rangle + \cos\left(G_{U}|\xi|t\right) |1,1,0\rangle \\ + e^{-i\Omega t} \sin\left(G_{U}|\xi|t\right) |1,0,1\rangle\Big),
\end{multline}
which is heavily modulated by the state of the gaseous molecular ensemble via $\xi$, and is analogous to the ones found in other quantum frequency conversion schemes~\cite{Clark2013}. 
\begin{figure}[t]
{\includegraphics[width=\linewidth]{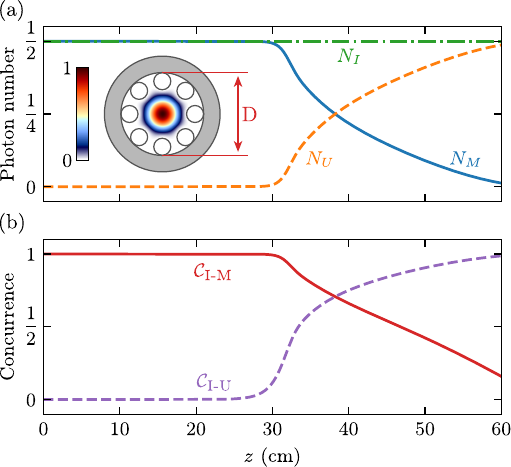}}
\caption{Evolution of the average photon numbers and concurrences along the fiber filled with 70 bar H$_2$ and pumped with 115 $\mu$J pulse energy. (a) Average photon numbers of the idler, $N_I$, mixing, $N_M$, and up-converted, $N_U$, frequency modes. The inset shows the simulated transverse intensity distribution (normalized to its maximum) of the fundamental core mode in a single-ring-type ARF similar to that used in Ref.~\onlinecite{Tyumenev2022} (see Table~I in the supplementary material). Here, D represents the larger internal diameter of the fiber. (b) Dynamics of the idler-mixing, $C_{\text{I-M}}$, and idler-up-converted, $C_{\text{I-U}}$, concurrences.}
\label{fig3}
\end{figure}
Therefore, in this framework, the resulting equations for the evolution of the mixing, $N_M$, and the up-converted, $N_U$, average photon numbers during molecular modulation are
\begin{eqnarray}
\label{photon_number_M} N_{M} &=& \frac{1}{2}\cos^{2}\left(G_{U}|\xi|t\right), \\
\label{photon_number_U} N_{U} &=& \frac{1}{2}\sin^{2}\left(G_{U}|\xi|t\right).
\end{eqnarray}
Meanwhile, we also study the dynamics of entanglement between the mixing and the up-converted frequencies through the concurrence~\cite{Wootters1998}, an entanglement monotone used for bipartite mixed states. This is an appropriate choice in this case since we have entanglement between the idler and the mixing subsystems, but also between the idler and up-converted. Furthermore, the concurrence completely characterizes the entanglement of formation~\cite{Hill1997} for a pair of two-level systems. In our case, we find that the idler-mixing and the idler-up-converted concurrences are
\begin{eqnarray}
C_{\text{I-M}} &=& \Big| \cos\left(G_{U}|\xi|t\right) \Big|, \\
C_{\text{I-U}} &=& \Big| \sin\left(G_{U}|\xi|t\right) \Big|.
\end{eqnarray}
Notice how entanglement transfer between mixing and up-converted modes is closely related to the evolution of the number of photons in each frequency mode, and thus higher conversion efficiencies lead to a more effective entanglement transfer. The time parameter $t$ here is considered to be related to the propagation distance $z$ inside the fiber used in Ref.~\onlinecite{Tyumenev2022}, $z\approx ct$. This is used to represent the evolution of the average photon numbers and the concurrences in Fig.~\ref{fig3}, where the explicit time dependence of $\alpha_{P}$ and $\alpha_{S}$, as shown in the supplementary material, has been considered. These coefficients are obtained by numerically solving the semiclassical Maxwell-Bloch equations of motion for the pump and Stokes electric fields\cite{Raymer1990,Chen2024}. These already include the phenomenological $T_2$ factor accounting for the temporal dephasing of the molecular coherence. In addition, in order to obtain the coupling strength, we have estimated the quantization volume based on the geometric properties of the waveguide such as the larger internal diameter D (see Fig.~\ref{fig3}) and the temporal length of the interaction.




\section{Results}

In Fig.~\ref{fig3}~(a), we show the average photon numbers for the mixing, up-converted, and idler frequencies by blue solid, orange dashed, and green dashed-dotted lines, respectively. In Fig.~\ref{fig3}~(b), we do the same for the idler-mixing and the idler-up-converted concurrences using red solid and purple dashed lines, respectively. We can observe that, until molecular coherence becomes significant at around the middle of the fiber, no substantial conversion dynamics occur; from that point on, the probability of frequency up-converting a mixing photon increases, and this is reflected in higher entanglement between idler and up-converted frequencies. Meanwhile, the idler-mixing concurrence decreases. Furthermore, notice that the crossings between the average photon numbers and the concurrences occur at exactly the same point, around $z\approx$ 38 cm, as a consequence of the close relation between the transfer of entanglement and the transfer of photon population. Note that, here, the maximum value for the average photon numbers is $1/2$ due to the type of photonic quantum state considered as input to the system. After a quarter of an oscillation, i.e. when $G_{U}|\xi|t = \pi/2$, we find the state $\left(|0,0,0\rangle + e^{-i\Omega t} |1,0,1\rangle\right)/\sqrt{2}$ for the idler, mixing, and up-converted, respectively. This state showcases that high efficiencies
in the transfer of a photon from the mixing to the up-converted mode can be theoretically attainable, as can be seen from Eqs.~\eqref{photon_number_M} and \eqref{photon_number_U}. Note that our predictions could be tested in future experiments, since efforts to characterize the concurrence of mixed states have already been made~\cite{Huang2009}, obtaining lower~\cite{Mintert2007} and upper bounds~\cite{Zhang2008} of this quantity.

\begin{figure}[h]
\centering
\includegraphics[width=\linewidth]{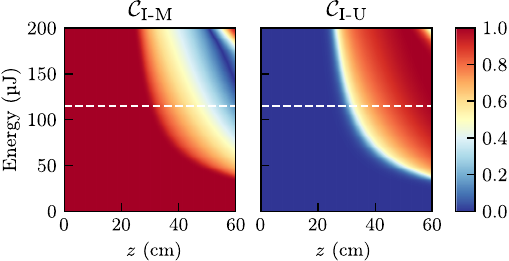}
\caption{Concurrence as a function of the initial pump beam energy and the propagation distance $\boldsymbol z$. The plots show the evolution of the idler-mixing $C_{\text{I-M}}$ and idler-up-converted $C_{\text{I-U}}$ concurrences at a pressure of 70 bar. The white dashed lines correspond to the evolution displayed in Fig.~\ref{fig3}.}
\label{fig4}
\end{figure}

A deeper analysis of the concurrence dynamics is presented in Fig.~\ref{fig4}, where the influence of the initial pump pulse energy is clearly shown. In general, the concurrence varies smoothly as the energy parameter is modified, indicating stable dynamics.  
In addition to this, Fig.~\ref{fig4} indicates that the transfer dynamics can be tuned to be produced at shorter $z$ values by increasing the initial pump pulse energy.

\section{Conclusion}


In conclusion, we have explored the use of molecular modulation triggered by SRS in gas-filled ARFs for frequency up-conversion of entangled photon pairs, showing how entanglement can be efficiently transferred to the frequency-converted counterpart. To do so, we employ a quantum Hamiltonian description capable of recovering the Maxwell-Bloch equations in the semiclassical limit, as shown in the supplementary material. With it, we were able to characterize the state of the molecules as the vibrational coherence is established, and to use it in the analysis of the molecular modulation of quantum light states injected in gas-filled ARFs. We derived simple expressions governing the evolution of entanglement in the system, demonstrating the close relationship with the evolution of the average photon numbers. 

We have found the single-mode treatment of both frequency and spatial modes to be satisfactory for the scope of this work, following the conditions of Ref.~\onlinecite{Tyumenev2022}. Nevertheless, a leap into a full multimode theory could be taken, for example, by following the quantum treatment in Ref.~\onlinecite{Wang2023}. This provides a description of the general effects of frequency translation in the shape of the second order time correlations of the experiment in Ref.~\onlinecite{Tyumenev2022} by considering the bandwidth in the quantum frequency mode distribution of the initial photonic state. Moreover, one can follow the works in Refs.~\onlinecite{Carman1970, Akhmanov1974, Raymer1979, Raymer1981}, which studied the role of the field bandwidths during the preparation of the Raman molecular coherence in the semiclassical regime. In addition, semiclassical treatments that deal with multiple spatial modes have also been considered~\cite{Bauerschmidt2015_2, Agrawal2007} following the lines of Ref.~\onlinecite{Raymer1990}.

As the experiments in this area are rapidly progressing, we believe that this framework will be a useful resource for the design of novel fiber-based quantum transduction strategies that could be fully integrated with existing fiber networks, thereby bringing the dream of the future quantum networks one step closer.

\section*{Supplementary Material}

In the supplementary material we provide more detailed calculations on the derivation of $H_{\text{eff}}$, the molecular state and the concurrences. The experimental parameters used at Ref.~\onlinecite{Tyumenev2022} and the subsequent time dependence of $\alpha_{P}$ and $\alpha_{S}$ can also be found.

\begin{acknowledgments}
We acknowledge financial support from HORIZON-CL4-2022-QUANTUM01-SGA project 101113946 OpenSuperQ-Plus100 of the EU Flagship on Quantum Technologies, the Spanish Ramón y Cajal Grant RYC-2020-030503-I, project Grants No. PID2021-125823NA-I00, PID2021-123131NA-I00, PID2021-122505OBC31 and TED2021-129959B-C21, funded by MICIU/AEI/10.13039/501100011033 and by “ERDF A way of making Europe”, by “ERDF Invest in your Future”, by the “European Union NextGenerationEU/PRTR” and "ESF+",  from the Basque Government through Grants No. IT1470-22 and IT1455-22 and ELKARTEK ($\mu$4Smart-KK-2023/00016, Ekohegaz II-KK-2023/00051, and KUBIT KK-2024/00105), and from the IKUR Strategy under the collaboration agreement between Ikerbasque Foundation and BCAM on behalf of the Department of Education of the Basque Government and the grant IKUR\_IKA\_23/04. ML acknowledges support from the predoctoral grant "Formación de Profesorado Universitario" FPU23/02350 from the Spanish Ministry of Science, Innovation and Universities (MICIU). This work has also been financially supported by the Ministry for Digital Transformation and the Civil Service of the Spanish Government through the QUANTUM ENIA project call – Quantum Spain project, and by the European Union through the Recovery, Transformation and Resilience Plan – NextGenerationEU within the framework of the Digital Spain 2026 Agenda.
\end{acknowledgments}

\section*{Data Availability Statement}

The data that support the findings of this study are available within the article and its supplementary material and from the corresponding author upon reasonable request.

\section*{Author Declarations}
\subsection*{\normalsize Conflict of Interest}

The authors have no conflicts to disclose.

\subsection*{\normalsize Author Contributions}

T.G.-R. and A.M. contributed equally to this work.\\

\textbf{Tasio Gonzalez-Raya:} Formal analysis (equal); Methodology (lead); Software (equal); Writing -- original draft (equal). \textbf{Arturo Mena:} Formal analysis (equal); Methodology (equal); Software (equal); Visualization (lead); Writing -- original draft (equal). \textbf{Miriam Lazo:} Methodology (supporting). \textbf{Luca Leggio:} Visualization (supporting); Writing -- review \& editing (equal). \textbf{David Novoa:} Conceptualization (equal); Funding Acquisition (equal); Supervision (equal); Writing -- original draft (supporting); Writing -- review \& editing (equal). \textbf{Mikel Sanz:} Conceptualization (lead); Methodology (equal); Funding Acquisition (equal); Supervision (equal); Writing -- review \& editing (equal);

\section*{References}

\clearpage
\widetext
\begin{center}
\textbf{\large Supplemental Material}
\end{center}
\setcounter{equation}{0}
\setcounter{figure}{0}
\setcounter{table}{0}
\setcounter{section}{0}
\makeatletter
\renewcommand{\theequation}{S\arabic{equation}}
\renewcommand{\thefigure}{S\arabic{figure}}
\renewcommand{\bibnumfmt}[1]{[S#1]}
\renewcommand{\citenumfont}[1]{S#1}

\section{Experimental parameters}\label{sm_1}
In this section, we present a list of the parameters extracted from the experiment in Ref.~\onlinecite{SM_Tyumenev2022} (see Table~\ref{table1}), which we have used to obtain the results presented in this manuscript. 
\begin{table}[h!] 
\centering 
\begin{tabular}{@{}llll@{}} \toprule[1.5pt] \multicolumn{4}{c}{List of experimental parameters} \\ \midrule[1.25pt] 
Parameter name & Symbol & Value & Units \\ \midrule[1.25pt] 
\rowcolor[gray]{.9} $\text{H}_{2}$ pressure & $P$ & 70 & bar \\ 
Temperature & $T$ & 298 & K \\ 
\rowcolor[gray]{.9}Pulse energy & $E_{\text{pulse}}$ & $1.15\times10^{-4}$ & J \\ 
Pulse temporal width & $T_{\text{width}}$ & $8.5432\times10^{-9}$ & s \\ 
\rowcolor[gray]{.9} Number of molecules & $N$ & $1.4925\times10^{18}$ & - \\
Raman shift & $\Omega/2\pi$ & $1.2457\times10^{14}$ & Hz \\
\rowcolor[gray]{.9} Phase relaxation time & $T_{2}$ & $9.6897\times10^{-11}$ & s \\
Damping rate& $\Gamma$ & $1.0320\times10^{10}$ & Hz \\
\rowcolor[gray]{.9} Fiber diameter & $D$ & $1.1\times10^{-4}$ & m \\
Fiber length & $L$ & 0.6 & m \\
\rowcolor[gray]{.9} Effective area of the $\text{LP}_{01}$ mode & $A_{01}$ & $1.4621\times10^{-9}$ & $\text{m}^{2}$ \\
Total area of the fiber & $A_{\text{fiber}}$ & $9.5033\times10^{-9}$ & $\text{m}^{2}$ \\
\rowcolor[gray]{.9} Effective volume of the fiber & $\mathcal{V}_{\text{fiber}}$ & $8.7723\times10^{-10}$ & $\text{m}^{3}$ \\
Quantization volume & $\mathcal{V}$ & $2.4340\times10^{-8}$ & $\text{m}^{3}$ \\
\rowcolor[gray]{.9} Pump wavelength & $\lambda_{P}$ & $1.064\times10^{-6}$ & m \\
Stokes wavelength & $\lambda_{S}$ & $1.9072\times10^{-6}$ & m \\
\rowcolor[gray]{.9} Anti-Stokes wavelength & $\lambda_{A}$ & $7.3781\times10^{-7}$ & m \\
Mixing wavelength & $\lambda_{M}$ & $1.425\times10^{-6}$ & m \\
\rowcolor[gray]{.9} Up-converted wavelength & $\lambda_{U}$ & $8.9503\times10^{-7}$ & m \\
Pump frequency & $\nu_{P}$ & $2.8176\times10^{14}$ & Hz \\
\rowcolor[gray]{.9} Stokes frequency & $\nu_{S}$ & $1.5719\times10^{14}$ & Hz \\
Anti-Stokes frequency & $\nu_{A}$ & $4.0633\times10^{14}$ & Hz \\
\rowcolor[gray]{.9} Mixing frequency & $\nu_{M}$ & $2.1038\times10^{14}$ & Hz \\
Up-converted frequency & $\nu_{U}$ & $3.3495\times10^{14}$ & Hz \\
\rowcolor[gray]{.9} Gain pump-Stokes & $\gamma_{p-s}$ & $9.7644\times10^{-12}$ & $\text{m W}^{-1}$ \\
Gain mixing-up-converted & $\gamma_{m-u}$ & $1.3233\times10^{-11}$ & $\text{m W}^{-1}$ \\
\rowcolor[gray]{.9} Coupling pump-Stokes & $\kappa_{1,p}$ & $-8.9518\times10^{-8}$ & $\text{m}^{2} \text{ C}^{2}\text{ J}^{-2} \text{ s}^{-1}$ \\
Coupling mixing-up-converted & $\kappa_{1,u}$ & $-9.0080\times10^{-8}$ & $\text{m}^{2} \text{ C}^{2}\text{ J}^{-2} \text{ s}^{-1}$ \\
\bottomrule[1.5pt] 
\end{tabular} 
\caption{Table containing experimental parameters from Ref.~\onlinecite{SM_Tyumenev2022} used in the simulations displayed in this letter.}
\label{table1} 
\end{table}

The number of molecules is computed using the ideal gas law,
\begin{equation}
N = \frac{P \mathcal{V}_{\text{fiber}}}{k_{\text{B}}T},
\end{equation}
where $k_{\text{B}}= 1.380649\times10^{-23} \text{ J K}^{-1}$ is the Boltzmann constant, $T$ is the temperature in Kelvin, $\mathcal{V}_{\text{fiber}}$ is the volume of the fiber in $\text{m}^{3}$, and $P$ is the pressure in Pa. Meanwhile, the total number of photons is calculated by dividing the energy of the pulse by the energy of a single pump photon, yielding $N_{\text{photons}} = 6.1598\times10^{14}$. The interaction strengths used in this letter are computed in terms of the couplings given in Table~\ref{table1}, as 
\begin{eqnarray}
G_{S} &=& -\kappa_{1,p} \frac{h\sqrt{\nu_{P}\nu_{S}}}{2\epsilon_{0}\mathcal{V}} = 2.8962\times10^{-8} \text{ Hz}, \\
G_{U} &=& -\kappa_{1,u} \frac{h\sqrt{\nu_{U}\nu_{M}}}{2\epsilon_{0}\mathcal{V}} = 3.6760\times10^{-8} \text{ Hz}.
\end{eqnarray}
Here, $\epsilon_{0}$ is the electric permittivity of the vacuum, and $\mathcal{V}$ is the quantization volume of the fields. The couplings are computed using the phenomenological relations~\cite{Bauerschmidt2016}
\begin{eqnarray}
\kappa_{1,p} &=& -\sqrt{\frac{2\gamma_{p-s}c^{2}\Gamma\epsilon_{0}^{2}}{N_{\rho}\hbar(\omega_{P}-\Omega)}}, \\
\kappa_{1,u} &=&  -\sqrt{\frac{2\gamma_{m-u}c^{2}\Gamma\epsilon_{0}^{2}}{N_{\rho}\hbar(\omega_{U}-\Omega)}}.
\end{eqnarray}

On the other hand, $\mathcal{V}$ is obtained by multiplying the total area of the fiber $A_{\text{fiber}}$ by the temporal width of the pulse $T_{\text{width}}$ and the speed of light $c$, i.e., $\mathcal{V}=c T_{\text{width}} A_{\text{fiber}}$. In this sense, $\mathcal{V}$ would follow the textbook definition of relevant volume that contains the energy of the radiation field, taking the volume integral of the electromagnetic energy density as reference. In order to provide a reasonable analytical expression for this volume, knowing that the real pulse has a finite duration, we define the temporal profile of the pulse as a piecewise function that is zero everywhere except in the relevant pulse region, where it would follow an amplitude distribution. In this regard, we consider the Bessel function $J_{0}$. Therefore, $T_{\text{width}}$ is estimated through the distance between the first zeros of $J_{0}$ that approximates the Gaussian distribution considered in Ref.~\onlinecite{SM_Tyumenev2022} for the pulse profile. The Bessel function used is determined by matching the integral of its square to the integral of the square of the previously mentioned Gaussian distribution in order to obtain the same total pulse energy while maintaining the same field amplitude factor. For the pulse transversal profile area, following the definition of volume containing the radiation, we considered $A_{\text{fiber}}$. This area is given by the inner diameter of the fiber at Ref.~\onlinecite{SM_Tyumenev2022}, excluding the capillaries. This is considered because, although hollow-core anti-resonant fibers present extremely low losses and provide tight modal confinement in the hollow region at the center of the multi-capillary microstructure, there is still some residual light intensity outside of the pulse's effective mode area. With this approach, the value obtained for $\mathcal{V}$ leads to reasonable dynamics given the evolution time, given that $G_{S}$ and $G_{U}$ depend on the quantization volume. 

\begin{figure}[h!]
{\includegraphics[width=85mm]{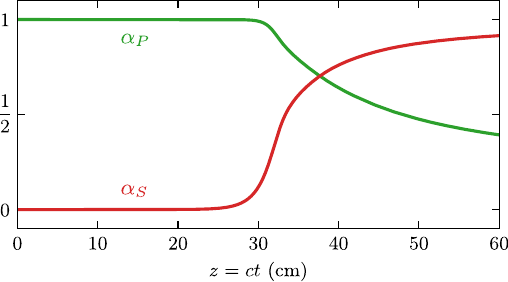}}
\caption{Evolution of the normalized amplitudes of the coherent states describing the pump and Stokes pulses during stimulated Raman scattering inside a hollow-core fiber filled with hydrogen gas as a function of the fiber length. We represent the normalized $\alpha_{P}$ and $\alpha_{S}$ in green and in red, respectively, against the length of the fiber. This figure illustrates how a coherent state is developed in the Stokes mode through the fiber, which leads to a depletion of the pump.  Meanwhile, coherence is being developed in the molecules, gaining relevance at around the half-length of the fiber, when the pump and Stokes amplitudes start to change. }
\label{fig_s1}
\end{figure}
In Fig.~\ref{fig_s1}, we present the evolution of $\alpha_{P}$ and $\alpha_{S}$, the amplitudes of the coherent states that characterize the pump and the Stokes pulses inside the fiber, as studied in Ref.~\onlinecite{SM_Tyumenev2022}. In this study, molecular coherence inside a hollow-core fiber filled with hydrogen gas is developed through a stimulated Raman scattering process. A pump tone is used to excite the molecules, producing an increase on the population of the corresponding Stokes frequency mode, as one can see in Fig.~\ref{fig_s1}. In green, one can see the normalized amplitude of the pump field while, in red, we have the normalized amplitude of Stokes photons. Most of all dynamics occur at the latter half of the fiber, where molecular coherence starts to develop.

\section{From original to effective Hamiltonian}\label{sm_2}
The Hamiltonian describing the process of Raman scattering can be split into an unperturbed part, $H_{0}$, and an interaction part, $V$. On the one hand, the unperturbed Hamiltonian can be expressed as
\begin{equation}
H_{0} = \hbar\omega_{0} |0\rangle\langle0| + \hbar\omega_{1} |1\rangle\langle1| + \hbar\sum_{i=2}^{\infty}\omega_{i} |i\rangle\langle i| + \hbar\sum_{l}\omega_{l}a^{\dagger}_{l}a_{l},
\end{equation}
where $\omega_{i}$ is the frequency associated to the transition between vibrational levels $|i\rangle$ and $|i+1\rangle$ of the molecule, whereas $\omega_{l}$ represents the frequency of mode $l$ of the electric field, with $l\in\{P,S,A\}$ labelling the pump, Stokes, and anti-Stokes frequencies, respectively. On the other hand, we derive the interaction Hamiltonian, assuming that the interaction is dipolar, from the term $-\vec{\bf \mu}\vec{\bf E}(z)$. Here, $\vec{\bf \mu}$ and $\vec{\bf E}(z)$ are the dipole and the electric field operators, respectively, and can be written as 
\begin{eqnarray}
\vec{\bf \mu} &=& \vec{e}\sum_{i,j}\mu_{i,j}|i\rangle\langle j|, \\
\vec{\bf E}(z) &=& i \vec{f} \sqrt{\frac{\hbar\omega}{2\epsilon_{0}\mathcal{V}}}\left( a e^{i\beta z} - a^{\dagger}e^{-i\beta z}\right).
\end{eqnarray} 
If these two are aligned, $\vec{e}\vec{f}=1$, we can write the interaction term as
\begin{equation}
-\vec{\bf \mu}\,\vec{\bf E}(z) = \sum_{i,j}g_{i,j} |i\rangle\langle j| \left( a e^{i\beta z} - a^{\dagger}e^{-i\beta z}\right),
\end{equation}
where we have defined 
\begin{equation}
g_{i,j} = -i\mu_{i,j}\sqrt{\frac{\hbar\omega}{2\epsilon_{0}\mathcal{V}}}.
\end{equation}
Notice that $g_{j,i} = -g_{i,j}^*$ has units of energy. Then, the interaction term for the Hamiltonian is given by
\begin{equation}
V = \sum_{i,j}\sum_{l} g_{i,j}^{l} |i\rangle\langle j| \left(a_{l}e^{i\beta_{l}z} - a_{l}^{\dagger}e^{-i\beta_{l}z}\right).
\end{equation}
 Now, we want to make an interaction picture transformation and eliminate the energy levels with $i>1$ from the Hamiltonian. We assume these levels are off resonance, and focus on the resonant transition between the ground state $|0\rangle$ and the first vibrational state $|1\rangle$. First, we go to an interaction picture with respect to $H_{0}$. For that, let us propose a splitting of the evolution operator into $U=U_{0}U_{t}$, such that the Schr\"{o}dinger equation reads
\begin{equation}
i\hbar\partial_{t}(U_{0}U_{t}) = (H_{0}+V)U_{0}U_{t}.
\end{equation}
If we expand the derivative, we arrive at
\begin{equation}
i\hbar U_{0}(\partial_{t}U_{t}) + (i\hbar\partial_{t}U_{0}-H_{0}U_{0})U_{t} = VU_{0}U_{t}.
\end{equation}
We have that $i\partial_{t}U_{0}-H_{0}U_{0}=0$, since it is the Schr\"{o}dinger equation for $H_{0}$, and we are left with
\begin{equation}
i\hbar\partial_{t}U_{t} =U_{0}^{\dagger}VU_{0}U_{t}.
\end{equation}
Assuming that $U_{0}=e^{-i H_{0} t/\hbar}$, because $H_{0}$ is not time dependent, the solution for $U_{t}$ is
\begin{eqnarray}
\nonumber U_{t} &=& \hat{T} e^{-\frac{i}{\hbar}\int_{0}^{t}\text{d}s \, e^{i H_{0}s/\hbar} V e^{-i H_{0}s/\hbar}} = \Id - \frac{i}{\hbar} \int_{0}^{t}\text{d}s \, e^{i H_{0}s/\hbar} V e^{-i H_{0}s/\hbar} \\
&+& \frac{(-i)^{2}}{\hbar^{2}} \int_{0}^{t}\text{d}s \, e^{i H_{0}s/\hbar} V e^{-i H_{0}s/\hbar} \int_{0}^{s}\text{d}s' \, e^{i H_{0}s'/\hbar} V e^{-i H_{0}s'/\hbar} + \ldots,
\end{eqnarray}
what is know as the Dyson series, with $\hat{T}$ being the time-ordering operator. Knowing the following formula,
\begin{equation}\label{BCH}
e^{A}Be^{-A} = \sum_{k=0}^{\infty} \frac{1}{k!}[A,[A,\ldots,[A,B]\ldots]]_{k},
\end{equation}
we compute the commutators of $H_{0}$ with $V$,
\begin{eqnarray}
[H_{0},V] &=& \hbar\sum_{i,j}\sum_{l} g_{i,j}^{l} |i\rangle\langle j| \left[ (\omega_{i}-\omega_{j}-\nu_{l})a_{l}e^{i\beta_{l}z} - (\omega_{i}-\omega_{j}+\nu_{l})a^{\dagger}_{l}e^{-i\beta_{l}z} \right], \\
\nonumber [H_{0},[H_{0},V]] &=& \hbar^{2}\sum_{i,j}\sum_{l} g_{i,j}^{l} |i\rangle\langle j| \left[ (\omega_{i}-\omega_{j}-\nu_{l})^{2} a_{l}e^{i\beta_{l}z} - (\omega_{i}-\omega_{j}+\nu_{l})^{2} a^{\dagger}_{l}e^{-i\beta_{l}z} \right].
\end{eqnarray}
We can infer from this that 
\begin{equation}
e^{iH_{0}t/\hbar}Ve^{-iH_{0}t/\hbar}\equiv V_{I}(t) = \sum_{i,j}\sum_{l} g_{i,j}^{l} |i\rangle\langle j| e^{i(\omega_{i}-\omega_{j})t}\left(a_{l}e^{i(\beta_{l}z - \omega_{l}t)} - a_{l}^{\dagger}e^{-i(\beta_{l}z - \omega_{l}t)}\right).
\end{equation} 

As it is often done in time-dependent perturbation theory, we expand to second order in $U_{t}$,
\begin{equation}
U_{t} = \Id - \frac{i}{\hbar} \int_{0}^{t}\text{d}s \, e^{i H_{0}s/\hbar} V e^{-i H_{0}s/\hbar} - \frac{1}{\hbar^{2}}\int_{0}^{t}\text{d}s \, e^{i H_{0}s/\hbar} V e^{-i H_{0}s/\hbar} \int_{0}^{s}\text{d}s' \, e^{i H_{0}s'/\hbar} V e^{-i H_{0}s'/\hbar},
\end{equation}
and compare it to the propagator given by the effective Hamiltonian we want to find,
\begin{equation}
U_{\text{eff}}(t) = \Id - \frac{i}{\hbar} \int_{0}^{t}\text{d}s \,  H_{\text{eff}}(s) + \ldots,
\end{equation}
which is normally kept at first order. We assume that the first-order term in $U_{t}$ can be adiabatically eliminated because $g_{i,j}^{l}\ll \omega_{i}-\omega_{j} \pm \omega_{l}$. Therefore, we are set to compare the terms $V_{I}(s)V_{I}(s')$ and $H_{\text{eff}}(s)$. We expand the latter and write
\begin{eqnarray}
\nonumber V_{I}(s)V_{I}(s') &=& \sum_{i,j,k}\sum_{l,m} g_{i,j}^{l}g_{j,k}^{m} |i\rangle\langle k| \Big( a^{\dagger}_{l}a^{\dagger}_{m} e^{-i(\beta_{l}+\beta_{m})z}e^{i(\omega_{i}-\omega_{j}+\omega_{l})s}e^{i(\omega_{j}-\omega_{k}+\omega_{m})s'} \\
\nonumber &-& a^{\dagger}_{l}a_{m} e^{-i(\beta_{l}-\beta_{m})z}e^{i(\omega_{i}-\omega_{j}+\omega_{l})s}e^{i(\omega_{j}-\omega_{k}-\omega_{m})s'} - a_{l}a^{\dagger}_{m} e^{i(\beta_{l}-\beta_{m})z}e^{i(\omega_{i}-\omega_{j}-\omega_{l})s}e^{i(\omega_{j}-\omega_{k}+\omega_{m})s'} \\
\nonumber &+& a_{l}a_{m} e^{i(\beta_{l}+\beta_{m})z}e^{i(\omega_{i}-\omega_{j}-\omega_{l})s}e^{i(\omega_{j}-\omega_{k}-\omega_{m})s'}\Big).
\end{eqnarray}
Since we want to identify $-iH_{\text{eff}}(s)/\hbar$ with $-V_{I}(s)\int_{0}^{s}\text{d}s' \, V_{I}(s')/\hbar^{2}$, we need to perform the integral over $s'$:
\begin{eqnarray}
\nonumber && V_{I}(s)\int_{0}^{s}\text{d}s' \, V_{I}(s') = -i\sum_{i,j,k}\sum_{l,m} g_{i,j}^{l}g_{j,k}^{m} |i\rangle\langle k| \Big[ \frac{a^{\dagger}_{l}a^{\dagger}_{m}e^{-i(\beta_{l}+\beta_{m})z}}{\omega_{j}-\omega_{k}+\omega_{m}} \Big( e^{i(\omega_{i}-\omega_{k}+\omega_{l}+\omega_{m})s}-e^{i(\omega_{i}-\omega_{j}+\omega_{l})s}\Big)  \\
\nonumber && - \frac{a^{\dagger}_{l}a_{m}e^{-i(\beta_{l}-\beta_{m})z}}{\omega_{j}-\omega_{k}-\omega_{m}} \Big( e^{i(\omega_{i}-\omega_{k}+\omega_{l}-\omega_{m})s}-e^{i(\omega_{i}-\omega_{j}+\omega_{l})s}\Big) - \frac{a_{l}a^{\dagger}_{m}e^{i(\beta_{l}-\beta_{m})z}}{\omega_{j}-\omega_{k}+\omega_{m}} \Big( e^{i(\omega_{i}-\omega_{k}-\omega_{l}+\omega_{m})s}-e^{i(\omega_{i}-\omega_{j}-\omega_{l})s}\Big) \\
&& + \frac{a_{l}a_{m}e^{i(\beta_{l}+\beta_{m})z}}{\omega_{j}-\omega_{k}-\omega_{m}} \Big( e^{i(\omega_{i}-\omega_{k}-\omega_{l}-\omega_{m})s}-e^{i(\omega_{i}-\omega_{j}-\omega_{l})s}\Big)\Big].
\end{eqnarray}
Basically, we are now going to neglect all rotating terms, in the approximation mentioned before. For this, we need to identify the resonant frequencies in the system. We define $\Omega_{i,j}\equiv \omega_{i}-\omega_{j}$, and identify $\Omega_{1,0}\equiv\Omega$ as the molecules vibrational transition frequency. Then, we can identify two resonances in the system, a Stokes and an anti-Stokes one, defined in relation to the pump frequency,
\begin{equation}
\Omega = \omega_{P}-\omega_{S} = \omega_{A}-\omega_{P}.
\end{equation}
Let us now compute the elements of $V_{I}(s)\int_{0}^{s}\text{d}s' \, V_{I}(s')$ in the basis of $\{|0\rangle, |1\rangle\}$,
\begin{eqnarray}
\nonumber && \langle0| V_{I}(s)\int_{0}^{s}\text{d}s' \, V_{I}(s')|0\rangle = i \sum_{k} |g_{0,k}^{l}|^{2} \sum_{l} \left( \frac{a_{l}^{\dagger}a_{l}}{\Omega_{k,0}-\omega_{l}} + \frac{a_{l}a_{l}^{\dagger}}{\Omega_{k,0}+\omega_{l}} \right), \\
\nonumber && \langle0| V_{I}(s)\int_{0}^{s}\text{d}s' \, V_{I}(s')|1\rangle = i \sum_{k} \left[ a^{\dagger}_{P}a_{S}e^{-i\Delta\beta  z}\left(\frac{g_{0,k}^{P}g_{k,1}^{S} }{\Omega_{k,1}-\omega_{S}} + \frac{g_{0,k}^{S}g_{k,1}^{P} }{\Omega_{k,1}+\omega_{P}} \right) + a^{\dagger}_{A}a_{P}e^{-i\Delta\beta' z}\left(\frac{g_{0,k}^{A}g_{k,1}^{P} }{\Omega_{k,1}-\omega_{P}} + \frac{g_{0,k}^{P}g_{k,1}^{A} }{\Omega_{k,1}+\omega_{A}} \right) \right], \\
\nonumber && \langle1| V_{I}(s)\int_{0}^{s}\text{d}s' \, V_{I}(s')|0\rangle = i \sum_{k} \left[ a_{P}a^{\dagger}_{S}e^{i\Delta\beta  z}\left(\frac{g_{1,k}^{S}g_{k,0}^{P} }{\Omega_{k,0}-\omega_{P}} + \frac{g_{1,k}^{P}g_{k,0}^{S}}{\Omega_{k,0}+\omega_{S}} \right) + a_{A}a^{\dagger}_{P}e^{i\Delta\beta' z}\left(\frac{g_{1,k}^{P}g_{k,0}^{A}}{\Omega_{k,0}-\omega_{A}} + \frac{g_{1,k}^{A}g_{k,0}^{P}}{\Omega_{k,0}+\omega_{P}} \right) \right], \\
\nonumber && \langle1| V_{I}(s)\int_{0}^{s}\text{d}s' \, V_{I}(s')|1\rangle = i \sum_{k} |g_{1,k}^{l}|^{2} \sum_{l} \left( \frac{a^{\dagger}_{l}a_{l}}{\Omega_{k,1}-\omega_{l}} + \frac{a_{l}a^{\dagger}_{l}}{\Omega_{k,1}+\omega_{l}} \right).
\end{eqnarray}
See that we have identified $\Delta\beta \equiv \beta_{P}-\beta_{S}$ and $\Delta\beta'\equiv \beta_{A}-\beta_{P}$. Then, we identify $H_{\text{eff}}$ as $-iV_{I}(s)\int_{0}^{s}\text{d}s' \, V_{I}(s')/\hbar$, and write
\begin{eqnarray}
\nonumber H_{\text{eff}}^{I}(t) &=& \hbar\sum_{l} a^{\dagger}_{l}a_{l} \left( \delta_{0,l}  |0\rangle\langle0| + \delta_{1,l} |1\rangle\langle1| \right) + \hbar\left( G_{0,1}^{S} a^{\dagger}_{P}a_{S} e^{-i\Delta\beta  z}e^{-i(\Omega-\omega_{P}+\omega_{S})t} + G_{0,1}^{A} a_{P}a^{\dagger}_{A} e^{-i\Delta\beta' z}e^{-i(\Omega-\omega_{A}+\omega_{P})t} \right) |0\rangle\langle1| \\
&+& \left( G_{1,0}^{S} a_{P}a^{\dagger}_{S} e^{i\Delta\beta  z}e^{i(\Omega-\omega_{P}+\omega_{S})t} + G_{1,0}^{A} a^{\dagger}_{P}a_{A} e^{i\Delta\beta' z}e^{i(\Omega-\omega_{A}+\omega_{P})t} \right) |1\rangle\langle0|.
\end{eqnarray}
Notice that this is defined in the interaction picture. In this Hamiltonian, we defined the following coefficients
\begin{eqnarray}
\delta_{0,l} &=& -\frac{1}{\hbar^{2}}\sum_{k}|g_{0,k}^{l}|^{2} \left( \frac{1}{\Omega_{k,0}-\omega_{l}} + \frac{1}{\Omega_{k,0}+\omega_{l}}\right), \\
\delta_{1,l} &=& -\frac{1}{\hbar^{2}}\sum_{k}|g_{1,k}^{l}|^{2} \left( \frac{1}{\Omega_{k,1}-\omega_{l}} + \frac{1}{\Omega_{k,1}+\omega_{l}}\right).
\end{eqnarray}
These are often referred to as dynamic Stark shifts. Furthermore, we have also defined
\begin{eqnarray}
G_{0,1}^{S} = \frac{1}{\hbar^{2}}\sum_{k} \left(\frac{g_{0,k}^{P}g_{k,1}^{S} }{\Omega_{k,1}-\omega_{S}} + \frac{g_{0,k}^{S}g_{k,1}^{P} }{\Omega_{k,1}+\omega_{P}} \right), \\
G_{0,1}^{A} = \frac{1}{\hbar^{2}}\sum_{k} \left(\frac{g_{0,k}^{A}g_{k,1}^{P} }{\Omega_{k,1}-\omega_{P}} + \frac{g_{0,k}^{P}g_{k,1}^{A} }{\Omega_{k,1}+\omega_{A}} \right), \\
G_{1,0}^{S} = \frac{1}{\hbar^{2}}\sum_{k} \left(\frac{g_{1,k}^{S}g_{k,0}^{P} }{\Omega_{k,0}-\omega_{P}} + \frac{g_{1,k}^{P}g_{k,0}^{S}}{\Omega_{k,0}+\omega_{S}} \right), \\
G_{1,0}^{A} = \frac{1}{\hbar^{2}}\sum_{k} \left(\frac{g_{1,k}^{P}g_{k,0}^{A}}{\Omega_{k,0}-\omega_{A}} + \frac{g_{1,k}^{A}g_{k,0}^{P}}{\Omega_{k,0}+\omega_{P}} \right).
\end{eqnarray}
Notice that here we can identify $G_{1,0}^{S}\equiv G_{S}$ and $G_{1,0}^{A}\equiv G_{A}$, such that $G_{0,1}^{S} = G_{S}^*$ and $G_{0,1}^{A} = G_{A}^*$, assuming that $g_{j,i}=g_{i,j}^*$. Then, we can write
\begin{eqnarray}
G_{S} = \frac{1}{\hbar^{2}}\sum_{k} \left(\frac{g_{1,k}^{S}g_{k,0}^{P} }{\Omega_{k,0}-\omega_{P}} + \frac{g_{1,k}^{P}g_{k,0}^{S}}{\Omega_{k,0}+\omega_{S}} \right), \\
G_{A} = \frac{1}{\hbar^{2}}\sum_{k} \left(\frac{g_{1,k}^{P}g_{k,0}^{A}}{\Omega_{k,0}-\omega_{A}} + \frac{g_{1,k}^{A}g_{k,0}^{P}}{\Omega_{k,0}+\omega_{P}} \right).
\end{eqnarray}
Let us point out some equivalences between frequencies,
\begin{eqnarray}
\nonumber \Omega_{k,1}-\omega_{S} &=& \Omega_{k,0}-\omega_{P}, \\
\nonumber \Omega_{k,1}+\omega_{P} &=& \Omega_{k,0}+\omega_{S}, \\
\nonumber \Omega_{k,1}-\omega_{P} &=& \Omega_{k,0}-\omega_{A}, \\
\nonumber \Omega_{k,1}+\omega_{A} &=& \Omega_{k,0}+\omega_{P}.
\end{eqnarray}
Let us now write the effective Hamiltonian in the Schr\"{o}dinger picture. We just have to cancel the exponentials, and recover the original Hamiltonian, $H_{0}$. 
\begin{eqnarray}
\nonumber H_{\text{eff}} &=& H_{0} + e^{-iH_{0}t/\hbar}H_{\text{eff}}^{I}(t)e^{iH_{0}t/\hbar} =  \\
\nonumber &=& \hbar\omega_{0}|0\rangle\langle0| + \hbar\omega_{1}|1\rangle\langle1| + \hbar\sum_{l} \omega_{l}a^{\dagger}_{l}a_{l} (|0\rangle\langle0|+|1\rangle\langle1|) + \hbar\sum_{l} a^{\dagger}_{l}a_{l} \left( \delta_{0,l}  |0\rangle\langle0| + \delta_{1,l} |1\rangle\langle1| \right) \\
&+& \hbar\left( G_{S} e^{i\Delta\beta  z}a_{P}a^{\dagger}_{S} + G_{A} e^{i\Delta\beta' z}a^{\dagger}_{P}a_{A}\right) |1\rangle\langle0| + \hbar\left( G_{S}^* e^{-i\Delta\beta  z} a^{\dagger}_{P}a_{S} + G_{A}^* e^{-i\Delta\beta' z} a_{P}a^{\dagger}_{A} \right) |0\rangle\langle1|.
\end{eqnarray}
We can rewrite this by replacing $|1\rangle\langle1|=(\Id+\sigma_{z})/2$, $|0\rangle\langle0|=(\Id-\sigma_{z})/2$, $|1\rangle\langle0|=\sigma^{+}$, and $|0\rangle\langle1|=\sigma^{-}$. We will neglect the constant term $\Id(\omega_{0}+\omega_{1})/2$, and define $\Delta_{l}^{\pm}=(\delta_{1,l}\pm\delta_{0,l})/2$. Finally, we obtain
\begin{eqnarray}
H_{\text{eff}} &=& \frac{\hbar\Omega}{2}\sigma_{z} + \hbar\sum_{l} \left( \omega_{l}+\Delta_{l}^{+} \right)a^{\dagger}_{l}a_{l} + \hbar\sum_{l} \Delta_{l}^{-}a^{\dagger}_{l}a_{l} \sigma_{z} + \hbar\left( G_{S}e^{i\Delta\beta  z} a_{P}a^{\dagger}_{S} + G_{A} e^{i\Delta\beta' z}a^{\dagger}_{P}a_{A}\right)\sigma^{+} \\
\nonumber &+& \hbar\left( G_{S}^* e^{-i\Delta\beta  z}a^{\dagger}_{P}a_{S} + e^{-i\Delta\beta' z} G_{A}^* a_{P}a^{\dagger}_{A} \right)\sigma^{-}.
\end{eqnarray}

In order to extend this to $N$ molecules, we need to replace $\sigma_{z}/2\rightarrow J_{z}$ and $\sigma^{\pm}\rightarrow J^{\pm}$, where we have identified
\begin{eqnarray}
J_{z} &=& \frac{1}{2}\sum_{i=1}^{N}\mathbb{1}_{1}\otimes\ldots\otimes\mathbb{1}_{i-1}\otimes\sigma_{i}^{z}\otimes\mathbb{1}_{i+1}\otimes\ldots\otimes\mathbb{1}_{N}, \\
J^{\pm} &=& \sum_{i=1}^{N}\mathbb{1}_{1}\otimes\ldots\otimes\mathbb{1}_{i-1}\otimes\sigma_{i}^{\pm}\otimes\mathbb{1}_{i+1}\otimes\ldots\otimes\mathbb{1}_{N}, 
\end{eqnarray}
with $[J_{z},J^{\pm}]=\pm J^{\pm}$, $[J^{+},J^{-}]=2J_{z}$, and the global spin states as $\left|{\bf \frac{N}{2}},m_{z}\right\rangle$, with $m_{z}\in\{-N/2,\ldots,N/2\}$. The global spin operators act on these states as follows,
\begin{eqnarray}
J_{z} \left | {\bf \frac{N}{2}},m_{z}\right\rangle &=& m_{z}\left | {\bf \frac{N}{2}},m_{z}\right\rangle, \\
J^{\pm} \left | {\bf \frac{N}{2}},m_{z}\right\rangle &=& \sqrt{\frac{N}{2}\left( \frac{N}{2}+1\right) - m_{z}\left(m_{z}\pm1\right) }\left | {\bf \frac{N}{2}},m_{z}\pm1\right\rangle.
\end{eqnarray}
With this, we can write the effective Hamiltonian describing the interaction with $N$ molecules,
\begin{eqnarray}\label{h_eff}
H_{\text{eff}} &=& \hbar\Omega J_{z} + \hbar\sum_{l} \left( \omega_{l}+\Delta_{l}^{+} \right)a^{\dagger}_{l}a_{l} + 2\hbar\sum_{l} \Delta_{l}^{-}a^{\dagger}_{l}a_{l} J_{z} + \hbar\left( G_{S}e^{i\Delta\beta  z} a_{P}a^{\dagger}_{S} + G_{A} e^{i\Delta\beta' z}a^{\dagger}_{P}a_{A}\right)J^{+} \\
\nonumber &+& \hbar\left( G_{S}^* e^{-i\Delta\beta  z}a^{\dagger}_{P}a_{S} + e^{-i\Delta\beta' z}G_{A}^* a_{P}a^{\dagger}_{A} \right)J^{-}.
\end{eqnarray}
As a final remark, we would like to point out that even though the adiabatic approximation performed through this second-order perturbation method is validated by semiclassical models~\cite{SM_Raymer1990}, which accurately describes the results of the experiment in Ref.~\onlinecite{SM_Tyumenev2022}, issues can arise when the assumptions of ideal resonance or short evolution times are not met, which can lead to a non-Hermitian effective Hamiltonian. In Ref.~\onlinecite{SM_Begzjav2022}, different
methods such as the Magnus expansion or the canonical transformation method are discussed,
which can help surpass these issues of non-Hermiticity.

\section{Semiclassical approximation for photon modes}\label{sm_3}
Using the Hamiltonian in Eq.~\eqref{h_eff} as a starting point, in the following we will consider that the anti-Stokes population is negligible, since the majority of the molecules remain in their ground state, and neglect the Stark shift terms. With these considerations, we will work with the following Hamiltonian, describing the pump-Stokes interactions. 
\begin{eqnarray}\label{H_eff_PS}
H_{\text{eff}} &=& \hbar\Omega J_{z} + \hbar\omega_{P}a^{\dagger}_{P}a_{P} + \hbar\omega_{S}a^{\dagger}_{S}a_{S} + \hbar \left(G_{S} e^{i\Delta\beta z}a_{P}a^{\dagger}_{S}J^{+} + G_{S}^{*}e^{-i\Delta\beta z}a^{\dagger}_{P}a_{S}J^{-}\right).
\end{eqnarray}
In the resonant case, $\Omega=\omega_P-\omega_S$, the interaction terms of the Hamiltonian, $\hbar \left(G_{S} e^{i\Delta\beta z}a_{P}a^{\dagger}_{S}J^{+} + G_{S}^{*}e^{-i\Delta\beta z}a^{\dagger}_{P}a_{S}J^{-}\right)$, commutes with the self energy terms, $\hbar\Omega J_{z} + \hbar\omega_{P}a^{\dagger}_{P}a_{P} + \hbar\omega_{S}a^{\dagger}_{S}a_{S}$, so we can study the evolution of the initial state under the Hamiltonian in the interaction picture with respect to these self energy terms, given in the equation below.
\begin{equation}\label{H_eff_PS_I}
H_{\text{eff}} ^{I}= \hbar \left(G_{S} e^{i\Delta\beta z}a_{P}a^{\dagger}_{S}J^{+} + G_{S}^{*}e^{-i\Delta\beta z}a^{\dagger}_{P}a_{S}J^{-}\right).
\end{equation}
In order to derive the molecular dynamics from this Hamiltonian, we will perform a semiclassical approximation in both photon modes. In the context of the experiment in Ref.~\onlinecite{SM_Tyumenev2022}, 115 $\mu$J pump pulses were used, which means that the initial pump state can be considered a coherent state with $\alpha_{P}\approx2.48\times10^{7}$. Then, since the annihilation of pump photons leads to the generation of intense laser radiation at the down-shifted Stokes frequency, we may also consider the Stokes field to be in a coherent state, with corresponding $\alpha_{S}$. With these considerations in mind, we define the following semiclassical Hamiltonian
\begin{equation}\label{H_alpha_I}
H_{\alpha}^{I}=\langle\alpha_P|\otimes\langle\alpha_S|\otimes\mathbb{1}H_{\text{eff}}^{I}|\alpha_P\rangle\otimes|\alpha_S\rangle\otimes\mathbb{1}=\hbar\left(G_{S} e^{i\Delta\beta  z} \alpha_{P}\alpha_{S}^{*}J^{+} + G_{S}^* e^{-i\Delta\beta z}\alpha_{P}^{*}\alpha_{S} J^{-}\right),
\end{equation}
which allows us to study the dynamics of the molecules. The Hamiltonian in Eq.~\eqref{H_alpha_I} is just the Hamiltonian obtained by replacing $a_P \rightarrow \alpha_P$, $a_P^\dagger \rightarrow \alpha_P^*$, $a_S \rightarrow \alpha_S$, $a_S^\dagger \rightarrow \alpha_S^*$ in in Eq.~\eqref{H_eff_PS_I}. In order to illustrate the validity of this approximation, it would be useful to compare the evolution of the state of the molecules given by the semiclassical Hamiltonian in Eq.~\eqref{H_alpha_I} $H_{\alpha}^{I}$, to that obtained with the original Hamiltonian neglecting the Stark shift and the self-energy terms, $H_{\text{eff}}^{I}$, acting on the full state of both the pump and the Stokes modes and the molecules after taking the partial trace over the bosonic modes. The density matrix of the state of the molecules under the evolution of the Hamiltonian in Eq.~\eqref{H_alpha_I} (Eq.~(4) in the manuscript) is given by
\begin{equation}
\rho_{\alpha} = e^{-\frac{it}{\hbar}H_{\alpha}^I}|\Psi\rangle\langle\Psi|e^{\frac{it}{\hbar}H_\alpha^I}=|\Psi\rangle\langle\Psi|-\frac{it}{\hbar}\left[H_{\alpha}^I,|\Psi\rangle\langle\Psi|\right] + \frac{1}{2}\left(\frac{it}{\hbar}\right)^2 \left[H_{\alpha}^I,\left[ H_{\alpha}^{I},|\Psi\rangle\langle\Psi|\right]\right]+\ldots
\end{equation}
where $|\Psi\rangle$ indicates the initial state of the molecules and we have expanded the exponential explicitly for the terms up to second order. On the other hand, the density matrix for the whole system is given by
\begin{equation}
\rho=e^{-\frac{it}{\hbar}H_{\text{eff}}^{I}}|\alpha_{P}\rangle\otimes|\alpha_{S}\rangle\otimes|\Psi\rangle\langle\alpha_{P}|\otimes\langle\alpha_{S}|\otimes\langle\Psi|e^{\frac{it}{\hbar}H_{\text{eff}}^{I}}.
\end{equation}
It is straightforward to see that if we expand the exponential and take partial trace over the photons states for the zeroth order term we would simply obtain $|\Psi\rangle\langle\Psi|$. We now examine the first order terms, $\rho^{(1)}$, that takes the following expression
\begin{equation}
\rho^{(1)} = -\frac{it}{\hbar}\left[ H_{\text{eff}}^{I},|\alpha_{P}\rangle\otimes|\alpha_{S}\rangle\otimes|\Psi\rangle\langle\alpha_{P}|\otimes\langle\alpha_{S}|\otimes\langle\Psi| \right].
\end{equation}
One can see that, by taking the partial trace over the bosonic modes, we find $\tr_{P,S}\rho^{(1)} = \rho_{\alpha}^{(1)}$. This means that, up to first order, the semiclassical approximation for the pump and Stokes fields is exact. Discrepancies arise when considering the second order terms, $\rho^{(2)}$, given by
\begin{equation}
\rho^{(2)} = \frac{1}{2}\left(\frac{it}{\hbar}\right)^2\left[H_{\text{eff}}^{I},\left[H_{\text{eff}}^{I},|\alpha_{P}\rangle\otimes|\alpha_{S}\rangle\otimes|\Psi\rangle\langle\alpha_{P}|\otimes\langle\alpha_{S}|\otimes\langle\Psi|\right]\right]
\end{equation}
By taking the partial trace over the bosonic modes, we find
\begin{eqnarray}
\nonumber \tr_{P,S}\rho^{(2)} &=& \frac{1}{2}\left(\frac{it}{\hbar}\right)^2 \left[H_{\alpha}^I,\left[ H_{\alpha}^{I},|\Psi\rangle\langle\Psi|\right]\right] \\
&+& \frac{1}{2}\left(\frac{it}{\hbar}\right)^2\hbar^2|G_s|^2 \Big[|\alpha_P|^{2} \left( J^{-}J^{+}|\Psi\rangle\langle\Psi| + |\Psi\rangle\langle\Psi|J^{-}J^{+} - 2 J^{+}|\Psi\rangle\langle\Psi|J^{-} \right) \\
\nonumber &+& |\alpha_S|^{2} \left( J^{+}J^{-}|\Psi\rangle\langle\Psi| + |\Psi\rangle\langle\Psi|J^{+}J^{-} - 2 J^{-}|\Psi\rangle\langle\Psi|J^{+} \right) \Big] 
\end{eqnarray}
See that we obtain the corresponding second-order term of the state evolved under the Hamiltonian in Eq.~\eqref{H_alpha_I}, $\rho_{\alpha}^{(2)}$, but we also find some new terms. In the semiclassical approximation, these are the first terms that we are neglecting. By comparing them to the largest coefficient appearing in the second order terms, which scale as $|\alpha_P|^2|\alpha_S|^2$, we can see that
\begin{eqnarray}
\nonumber\frac{\tr_{P,S}\rho^{(2)}-\rho_\alpha^{(2)}}{\hbar^2|G_s|^2|\alpha_P|^2|\alpha_S|^2}&=& \frac{1}{2}\left(\frac{it}{\hbar}\right)^{2}\bigg(-\frac{2}{|\alpha_S|^2}J^-|\Psi\rangle\langle\Psi|J^+-\frac{2}{|\alpha_P|^2}J^+|\Psi\rangle\langle\Psi|J^-+\frac{1}{|\alpha_S|^2}|\Psi\rangle\langle\Psi|J^-J^+\\
&+&\frac{1}{|\alpha_P|^2}|\Psi\rangle\langle\Psi|J^+J^-+\frac{1}{|\alpha_S|^2}J^-J^+|\Psi\rangle\langle\Psi|+\frac{1}{|\alpha_P|^2}J^+J^-|\Psi\rangle\langle\Psi|\bigg).
\end{eqnarray}
Since the order of magnitude of $\alpha_P,\ \alpha_S$ is $\sim 10^7$, these terms introduce a factor of $\sim 10^{-14}$ with respect to the leading terms, $\rho_{\alpha}^{(2)}$, and are therefore negligible. Then, the evolution of the molecular state is well-described by the propagator associated to the Hamiltonian $H_{\alpha}^{I}$ in Eq.~\eqref{H_alpha_I}.

\section{Spin Coherent State}\label{sm_4}
The pump-Stokes N-molecule effective Hamiltonian in the Schr\"{o}dinger picture is
\begin{equation}
H_{\text{eff}} = \hbar\Omega J_{z} + \hbar\omega_{P}a_{P}^{\dagger}a_{P}+\hbar\omega_{S}a_{S}^{\dagger}a_{S}+ \hbar\left(G_{S} e^{i\Delta\beta  z}a_{P}a_{S}^{\dagger}J^{+} + G_{S}^* e^{-i\Delta\beta  z}a_{P}^{\dagger}a_{S} J^{-}\right),
\end{equation}
where we have neglected the Stark shifts $\Delta_{l}^{\pm}$. We now transform $H_{\text{eff}}$ to an interaction picture with respect to $\Omega J_{z} + \omega_{P}a_{P}^{\dagger}a_{P}+\omega_{S}a_{S}^{\dagger}a_{S}$ in order to avoid oscillations with frequency $\Omega$ of our observables. There, we find 
\begin{equation}
H_\text{eff}^{I} = \hbar\left(G_{S} e^{i[(\Omega -\omega_{P}+\omega_{S})t + \Delta\beta  z]} a_{P}a_{S}^{\dagger}J^{+} + G_{S}^* e^{-i[(\Omega -\omega_{P}+\omega_{S})t + \Delta\beta  z]}a_{P}^{\dagger}a_{S} J^{-}\right).
\end{equation}
Notice that, in resonance conditions, $\Omega-\omega_{P}+\omega_{S}=0$, and there is no explicit time dependence in the Hamiltonian. Finally, we perform the semiclassical approximation described on Sec.\ref{sm_3} on the pump and Stokes modes $a_{l}\rightarrow\alpha_{l}$, and obtain
\begin{equation}
H_{\alpha}^{I} = \hbar\left(G_{S} e^{i\Delta\beta  z} \alpha_{P}\alpha_{S}^{*}J^{+} + G_{S}^* e^{-i\Delta\beta z}\alpha_{P}^{*}\alpha_{S} J^{-}\right).
\end{equation}
We will write the propagator associated with this Hamiltonian as
\begin{equation}
e^{-it H_{\alpha}^{I}/\hbar} = e^{-i t \left( \gamma J^{+} + \gamma^* J^{-}\right)}
\end{equation}
by defining $\gamma = G_{S}\alpha_{P}\alpha_{S}^* e^{i\Delta\beta z}$. Recall that the commutation relations for the global spin operators are
\begin{eqnarray}
\left[J_{z},J^{\pm}\right] &=& \pm J^{\pm}, \\
\left[J^{+},J^{-}\right] &=& 2J_{z}.
\end{eqnarray}
We assume that, since these commutators define a Lie algebra, there must be a relation such that~\cite{Truax1985}
\begin{equation}
e^{-i t\left( \gamma J^{+} + \gamma^* J^{-}\right)} = e^{s(t)J^{+}}e^{s_{0}(t)J_{z}}e^{s_{1}(t)J^{-}},
\end{equation}
with $s(0)=s_{0}(0)=s_{1}(0)=0$. By differentiating by $t$ on both sides, and then multiplying by the inverse of the right-hand side, we find
\begin{equation}
-i\left(\gamma J^{+} + \gamma^* J^{-} \right) = \dot{s}(t)J^{+} + \dot{s}_{0}(t)e^{s(t)J^{+}}J_{z}e^{-s(t)J^{+}} + \dot{s}_{1}(t)e^{s(t)J^{+}}e^{s_{0}(t)J_{z}}J^{-}e^{-s_{0}(t)J_{z}}e^{-s(t)J^{+}}.
\end{equation}
Using the relation in Eq.~\eqref{BCH}, we arrive at
\begin{equation}
-i\left(\gamma J^{+} + \gamma^* J^{-}\right) = \dot{s}(t)J^{+} + \dot{s}_{0}(t)\left( J_{z} - s(t)J^{+}\right) + \dot{s}_{1}(t)e^{-s_{0}(t)}\left( J^{-} + 2s(t)J_{z} - s^{2}(t)J^{+}\right),
\end{equation}
from where we find the set of differential equations
\begin{eqnarray}
&& \dot{s}(t) - \dot{s}_{0}(t)s(t) - \dot{s}_{1}(t)e^{-s_{0}(t)}s^{2}(t) = -i\gamma, \\
&& \dot{s}_{1}(t)e^{-s_{0}(t)} = -i\gamma^* , \\ 
&& \dot{s}_{0}(t) + 2 \dot{s}_{1}(t)e^{-s_{0}(t)}s(t) = 0.
\end{eqnarray}
Notice that we can combine these to find
\begin{eqnarray}
&& \dot{s}_{1}(t)e^{-s_{0}(t)} = -i\gamma^*, \\ 
&& \dot{s}_{0}(t) =  2i\gamma^*s(t).
\end{eqnarray}
and then we can isolate the equation for $s$,
\begin{equation}
\dot{s}(t) -i\gamma^* s^{2}(t) = -i\gamma.
\end{equation}
Solving this equation, we find
\begin{equation}
s(t) = -i\sqrt{\frac{\gamma}{\gamma^*}}\tan|\gamma|t.
\end{equation}
Using this to solve the remaining equations, we obtain
\begin{eqnarray}
s_{1}(t) &=& -i\sqrt{\frac{\gamma^*}{\gamma}}\tan|\gamma|t, \\
s_{0}(t) &=& -2\log\left(\cos|\gamma|t\right).
\end{eqnarray}
Assuming that $\alpha_{P}$, $\alpha_{S}$, and $G_{S}$ are real, we can simplify these as
\begin{eqnarray}
s(t) &=& e^{i\left(\Delta\beta  z -\frac{\pi}{2}\right)}\tan\left(G_{S}\alpha_{P}\alpha_{S} t\right), \\[0.5ex]
s_{1}(t) &=& e^{-i\left(\Delta\beta  z + \frac{\pi}{2}\right)}\tan\left(G_{S}\alpha_{P}\alpha_{S} t\right), \\[1ex]
s_{0}(t) &=& -2\log\left[ \cos\left(G_{S}\alpha_{P}\alpha_{S} t\right)\right].
\end{eqnarray}
Let us use this result to split the propagator $e^{-iH_{\alpha}^{I}t/\hbar}$, and obtain the state of the molecules at time $t$, assuming that they are initially in a ground state. In the interaction picture, the ground state of the molecules becomes
\begin{equation}
e^{-i\Omega t \frac{N}{2}}\left|{\bf \frac{N}{2}},-\frac{N}{2}\right\rangle.
\end{equation}
Then, we have that
\begin{equation}
|\psi_{I}(t)\rangle = e^{-i\Omega t \frac{N}{2}} e^{-iH_{\alpha}^{I}t/\hbar} \left|{\bf \frac{N}{2}},-\frac{N}{2}\right\rangle = e^{-i\Omega t \frac{N}{2}}e^{s(t)J^{+}}e^{s_{0}(t)J_{z}}e^{s_{1}(t)J^{-}} \left|{\bf \frac{N}{2}},-\frac{N}{2}\right\rangle,
\end{equation}
such that $e^{s_{1}(t)J^{-}}\left|{\bf \frac{N}{2}},-\frac{N}{2}\right\rangle=\left|{\bf \frac{N}{2}},-\frac{N}{2}\right\rangle$ and $e^{s_{0}(t)J_{z}}\left|{\bf \frac{N}{2}},-\frac{N}{2}\right\rangle=e^{-s_{0}(t)\frac{N}{2}}\left|{\bf \frac{N}{2}},-\frac{N}{2}\right\rangle$. We then find that
\begin{equation}
|\psi_{I}(t)\rangle = e^{-i\Omega t \frac{N}{2}} e^{-s_{0}(t)\frac{N}{2}}e^{s(t)J^{+}}\left|{\bf \frac{N}{2}},-\frac{N}{2}\right\rangle = e^{-i\Omega t \frac{N}{2}}e^{-s_{0}(t)\frac{N}{2}}\sum_{n=0}^{N}\frac{s^{n}(t)}{n!}(J^{+})^{n}\left|{\bf \frac{N}{2}},-\frac{N}{2}\right\rangle.
\end{equation}
If we compute explicitly the action of $J^{+}$, we find that
\begin{equation}
(J^{+})^{n}\left|{\bf \frac{N}{2}},-\frac{N}{2}\right\rangle = \sqrt{\frac{N! n!}{(N-n)!}}\left|{\bf \frac{N}{2}},-\frac{N}{2}+n\right\rangle,
\end{equation}
and we can write the final state of the molecules as
\begin{equation}
|\psi_{I}(t)\rangle = e^{-i\Omega t \frac{N}{2}}e^{-s_{0}(t)\frac{N}{2}}\sum_{n=0}^{N}\begin{pmatrix} N \\ n\end{pmatrix}^{1/2} s^{n}(t)\left|{\bf \frac{N}{2}},-\frac{N}{2}+n\right\rangle.
\end{equation}
This state should be normalized, so let us check if
\begin{equation}
\langle\psi_{I}(t)|\psi_{I}(t)\rangle = e^{-N s_{0}(t)}\left( 1 + |s(t)|^{2}\right)^{N} = 1.
\end{equation}
After some math, we can see that $s_{0}$ and $s$ are related through
\begin{equation}
e^{-s_{0}(t)} = \cos^{2}(G_{S}\alpha_{P}\alpha_{S} t) = \frac{1}{1+|s(t)|^{2}}.
\end{equation}
Therefore, we could write our state at time $t$ as 
\begin{equation}
|\psi_{I}(t)\rangle = \left(\frac{e^{-i\Omega t}}{1+|s(t)|^{2}}\right)^{\frac{N}{2}}\sum_{n=0}^{N}\begin{pmatrix} N \\ n\end{pmatrix}^{1/2} s^{n}(t)\left|{\bf \frac{N}{2}},-\frac{N}{2}+n\right\rangle.
\end{equation}
Note that this state is expressed in the interaction picture of $\Omega J_{z}+\omega_{P}a_{P}^{\dagger}a_{P}+\omega_{S}a_{S}^{\dagger}a_{S}$; if we return to the Schr\"{o}dinger picture, we find
\begin{equation}
|\psi(t)\rangle = \left(1+|s(t)|^{2}\right)^{-\frac{N}{2}}\sum_{n=0}^{N}\begin{pmatrix} N \\ n\end{pmatrix}^{1/2} \left(s(t) e^{-i\Omega t}\right)^{n}\left|{\bf \frac{N}{2}},-\frac{N}{2}+n\right\rangle.
\end{equation}

\section{Concurrence}\label{sm_5}
\begin{figure}[t]
    \centering
    \includegraphics[width=0.5\linewidth]{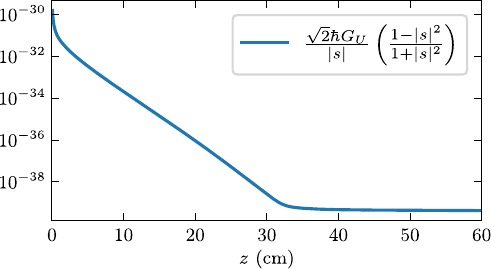}
    \caption{Plot of the norm of the Hamiltonian difference $||H_{\xi}^{I}-H_{\xi'}^{I}||$ in logarithmic scale in the limit $N\rightarrow\infty$ as a function of the propagation distance $z$. The value of $|s|$ is computed with the parameters used for Fig.~3 of the main text.}
    \label{fig_s2}
\end{figure}

After vibrational molecular coherence has been generated in the molecule ensemble via stimulated Raman scattering, we will consider that one frequency of an entangled state has been sent through the fiber to change its frequency. Unlike usual stimulated Raman scattering, in which the pump beam that is scattered to the different Raman frequency sidebands needs to have a power higher than a minimum power threshold, this is a thresholdless process which can be applied to single photons. We consider $(|0,0,0\rangle+|1,1,0\rangle)/\sqrt{2}$ as the initial state, where the states in tensor product, from left to right, indicate the photon number population of the idler frequency, kept outside of the fiber, the mixing frequency, which goes initially through the fiber, and the converted frequency, initially in a vacuum state. While the phase-difference between the mixing and the converted signals has to be equal to the phase of the molecular coherence wave, the jump in frequency is given by the Raman shift. Therefore, even though this is not, properly speaking, a Raman process, we can use the same interaction Hamiltonian to describe it. Furthermore, we will focus on the interaction between the pump and the anti-Stokes, since we are looking at frequency up-conversion. Then, we start from the Hamiltonian
\begin{equation}
H_{\text{eff}} = \hbar\Omega J_{z} + \hbar\omega_{M}a_{M}^{\dagger}a_{M} + \hbar \omega_{U}a_{U}^{\dagger}a_{U} + \hbar G_{U}\left( e^{-i(\beta_{M}-\beta_{U})z}a_{M}^{\dagger} a_{U}J^{+}  + e^{i(\beta_{M}-\beta_{U})z}a_{M} a_{U}^{\dagger}J^{-}\right),
\end{equation}
where we have introduced $G_{U}$ as the coupling strength between the mixing, the up-converted and the molecules, which can be defined as
\begin{equation}
G_{U} = \frac{1}{\hbar^{2}}\sum_{k} \left(\frac{g_{1,k}^{M}g_{k,0}^{U}}{\Omega_{k,0}-\omega_{U}} + \frac{g_{1,k}^{U}g_{k,0}^{M}}{\Omega_{k,0}+\omega_{M}} \right).
\end{equation}
Again, we will go to an interaction picture, this time with respect to $\Omega J_{z} + \omega_{M}a_{M}^{\dagger}a_{M} + \omega_{U}a_{U}^{\dagger}a_{U}$, resulting in 
\begin{equation}
H_\text{eff}^{I} = \hbar G_{U}\left( e^{i\big[(\Omega + \omega_{M}-\omega_{U})t-(\beta_{M}-\beta_{U})z\big]}a_{M}^{\dagger} a_{U}J^{+}  + e^{-i\big[(\Omega + \omega_{M}-\omega_{U})t-(\beta_{M}-\beta_{U})z\big]}a_{M} a_{U}^{\dagger}J^{-}\right).
\end{equation}
Meanwhile, the molecules will be in a coherent state and, analogous to what we previously did for the bosonic modes, we will perform a semiclassical approximation. This time, we will assume the state of the molecules will not change when we introduce a single excitation into the fiber, and replace the global spin operators by their expectation values over the spin coherent state derived in the previous section. This way, we can define the variable
\begin{equation}
\xi = \langle s|J^{-}|s\rangle = \frac{N s}{1+|s|^{2}}=\frac{N}{2}e^{i\left(\Delta\beta z - \frac{\pi}{2}\right)}\sin\left(2G_{S}\alpha_{P}\alpha_{S} t\right),
\end{equation}
in the interaction picture. This enters into our Hamiltonian as follows,
\begin{equation}
H^{I}_{\xi} = \hbar G_{U}\left( \xi^{*}e^{i\big[(\Omega + \omega_{M}-\omega_{U})t-(\beta_{M}-\beta_{U})z\big]}a_{M}^{\dagger} a_{U} + \xi e^{-i\big[(\Omega + \omega_{M}-\omega_{U})t-(\beta_{M}-\beta_{U})z\big]}a_{M} a_{U}^{\dagger}\right).
\end{equation}
Under phase matching and resonance conditions, we have $\beta_{U}-\beta_{M}=\Delta\beta\equiv\beta_{P}-\beta_{S}$ and $\omega_{U}-\omega_{M}=\Omega\equiv\omega_{P}-\omega_{S}$, respectively. Note that we do not require $\omega_l=\omega_m$ or $\beta_l=\beta_m$ with $l\in\{M,U\}$ and $m\in\{P,S\}$. Assuming these phase matching and resonance conditions are satisfied, this Hamiltonian simplifies to
\begin{equation}\label{H_eff_molecules}
H^{I}_{\xi} = \hbar G_{U}\left( \xi^{*}e^{i\Delta\beta z}a_{M}^{\dagger} a_{U} + \xi e^{-i\Delta\beta z}a_{M} a_{U}^{\dagger}\right).
\end{equation}
To test the validity of the semiclassical approximation on the global spin operators, we first compare the molecular states $|s\rangle$ and $|s'\rangle$, which describe spin coherent states that differ in a single molecular excitation, such that $|s'\rangle=J^{+}|s\rangle/\sqrt{\langle s|J^{-}J^{+}|s\rangle}$.  We use the trace distance to compare these states, finding 
\begin{equation}
T\left(|s\rangle,|s'\rangle\right) = \frac{1}{\sqrt{1+N|s|^{2}}}.
\end{equation}
One can see that, as the number of molecules increases, this distance goes to zero. This suggests that, in the single-excitation subspace, the molecular state will not change significantly, provided that the number of molecules is large ($\sim 10^{18}$ in Ref.~\onlinecite{SM_Tyumenev2022}). To go a bit further, we would like to compare the Hamiltonian in Eq.~\eqref{H_eff_molecules} to $H_{\xi'}^{I}$, with $\xi'=\langle s'|J^{-}|s'\rangle$, in order to comprehend the magnitude of the difference between the two in the single-excitation subspace. The norm of this difference is
\begin{equation}
\left|\left| H_{\xi}^{I}-H_{\xi'}^{I} \right|\right| = \sqrt{2} \hbar G_{U} \left|\xi-\xi'\right| = \sqrt{2} \hbar G_{U} |s| \left| \frac{2}{1+|s|^{2}} - \frac{N}{1+N|s|^{2}}\right|.
\end{equation}
In the limit $N\rightarrow\infty$, we find
\begin{equation}
\lim_{N\rightarrow\infty}\left|\left| H_{\xi}^{I}-H_{\xi'}^{I} \right|\right| = \frac{\sqrt{2} \hbar G_{U}}{|s|}\left| \frac{1-|s|^{2}}{1+|s|^{2}}\right|,
\end{equation}
which, as we can see at Fig.~\ref{fig_s2}, is very close to zero with the parametric values used in this work. In view of these results, we conclude that the Hamiltonian in Eq.~\eqref{H_eff_molecules} is adequate for working in the single excitation subspace. 


Now, we need to obtain the action of this Hamiltonian onto the initial photonic state of the system, which will be $(|0,0,0\rangle+|1,1,0\rangle)/\sqrt{2}$. But first, we need to transform this state into the interaction picture, 
\begin{equation}
\frac{1}{\sqrt{2}}\left( |0,0,0\rangle+e^{i\omega_{M}t}|1,1,0\rangle\right).
\end{equation}
We can easily see that $H_{\xi}^{I}|0,0,0\rangle = 0$, as well as
\begin{eqnarray}
\nonumber H_{\xi}^{I}|1,1,0\rangle &=& \hbar G_{U}\xi e^{-i\Delta\beta z}|1,0,1\rangle, \\[0.5ex]
\nonumber H_{\xi}^{I}|1,0,1\rangle &=& \hbar G_{U}\xi^* e^{i\Delta\beta z}|1,1,0\rangle.
\end{eqnarray}
Since this forms a closed subspace for a single excitation, we can diagonalize the Hamiltonian in this subspace, finding that the energies are $E_{\xi}^{\pm} = \pm \hbar G_{U}|\xi|$. Therefore, we propose the eigenstates to be of the form $|\phi_{\xi}^{\pm}\rangle = a|1,1,0\rangle \pm b|1,0,1\rangle$. If we insert this into the eigenvalue equation, we find that the eigenstates are
\begin{equation}
|\phi_{\xi}^{\pm}\rangle = \frac{1}{\sqrt{2}}\left( |1,1,0\rangle \pm e^{-i\Delta\beta z}\sqrt{\frac{\xi}{\xi^*}} |1,0,1\rangle\right).
\end{equation}
With this, we can find the evolution of the initial state,
\begin{equation}
\frac{1}{\sqrt{2}}\left( |0,0,0\rangle + e^{i\omega_{M}t}\cos(G_{U}|\xi|t)|1,1,0\rangle -i e^{i(\omega_{M} t-\Delta\beta z)}\sqrt{\frac{\xi}{\xi^*}}\sin(G_{U}|\xi|t)|1,0,1\rangle\right).
\end{equation}
We can further simplify this by noticing that $\xi/\xi^* = e^{2i\left(\Delta\beta  z-\frac{\pi}{2}\right)}$, such that the final state becomes
\begin{equation}
\frac{1}{\sqrt{2}}\left( |0,0,0\rangle + e^{i\omega_{M}t}\cos(G_{U}|\xi|t)|1,1,0\rangle + e^{i\omega_{M} t}\sin(G_{U}|\xi|t)|1,0,1\rangle\right).
\end{equation}
Back in the Schr\"{o}dinger picture, this state is
\begin{equation}
\frac{1}{\sqrt{2}}\left( |0,0,0\rangle + \cos(G_{U}|\xi|t)|1,1,0\rangle + e^{-i\Omega t}\sin(G_{U}|\xi|t)|1,0,1\rangle\right).
\end{equation}
The global density matrix is given by 
\begin{eqnarray}
\nonumber \rho &=& \frac{1}{2}\left( |0,0,0\rangle + \cos(G_{U}|\xi|t)|1,1,0\rangle + e^{-i\Omega t}\sin(G_{U}|\xi|t)|1,0,1\rangle\right)\times\\
&& \left( \langle0,0,0| + \cos(G_{U}|\xi|t)\langle1,1,0| + e^{i\Omega t}\sin(G_{U}|\xi|t)\langle1,0,1|\right).
\end{eqnarray}
We obtain the idler-mixing density matrix by doing the partial trace of $\rho$ with respect to the up-converted subsystem,
\begin{equation}
\rho_{I,M} = \frac{1}{2}\Big[  \big(|0,0\rangle + \cos(G_{U}|\xi|t)|1,1\rangle\big)\big( \langle0,0| + \cos(G_{U}|\xi|t)\langle1,1|\big)+ \sin^{2}(G_{U}|\xi|t)|1,0\rangle\langle1,0|\Big],
\end{equation}
and the idler-up-converted density matrix is obtained similarly from $\rho$ by tracing out the mixing frequency subsystem, 
\begin{equation}
\rho_{I,U} = \frac{1}{2}\Big[ \left(|0,0\rangle + e^{-i\Omega t}\sin(G_{U}|\xi|t)|1,1\rangle\right)\left( \langle0,0| + e^{i\Omega t}\sin(G_{U}|\xi|t)\langle1,1|\right)+ \cos^{2}(G_{U}|\xi|t)|1,0\rangle\langle1,0|\Big].
\end{equation}
If we write these in matrix form, we have
\begin{eqnarray}
\rho_{I,M} &=& \frac{1}{2}\begin{pmatrix} \cos^{2}G_{U}|\xi|t & 0 & 0 & \cos G_{U}|\xi|t \\ 0 & \sin^{2}G_{U}|\xi|t & 0 & 0 \\ 0 & 0 & 0 & 0 \\ \cos G_{U}|\xi|t & 0 & 0 & 1 \end{pmatrix}, \\[1ex]
\rho_{I,U} &=& \frac{1}{2}\begin{pmatrix} \sin^{2}G_{U}|\xi|t & 0 & 0 & e^{-i\Omega t}\sin G_{U}|\xi|t \\ 0 & \cos^{2}G_{U}|\xi|t & 0 & 0 \\ 0 & 0 & 0 & 0 \\ e^{i\Omega t}\sin G_{U} |\xi|t & 0 & 0 & 1 \end{pmatrix}.
\end{eqnarray}
Notice that we can also write these as
\begin{eqnarray}
\rho_{I,M} &=& \begin{pmatrix} x & 0 & 0 & w \\ 0 & \frac{1}{2} - x & 0 & 0 \\ 0 & 0 & 0 & 0 \\ w^* & 0 & 0 & \frac{1}{2} \end{pmatrix}, \\[1ex]
\rho_{I,U} &=& \begin{pmatrix} \frac{1}{2} - x & 0 & 0 & y \\ 0 & x & 0 & 0 \\ 0 & 0 & 0 & 0 \\ y^* & 0 & 0 & \frac{1}{2} \end{pmatrix}.
\end{eqnarray}
with $x = \frac{1}{2}\cos^{2}G_{U}|\xi|t$, $w =  \frac{1}{2}\cos G_{U}|\xi|t$, and $y = \frac{1}{2}e^{-i\Omega t}\sin G_{U}|\xi|t$. 

The entanglement measure we will use here is the concurrence, a bipartite entanglement metric used for two-qubit states, generally convenient for mixed states. In our case, we will look at bipartite entangled states that arise from a three-mode entanglement after tracing one mode in each case. The concurrence is defined as 
\begin{equation}
C = \max\{0,\lambda_{1}-\lambda_{2}-\lambda_{3}-\lambda_{4}\},
\end{equation}
where the $\lambda_{i}$ are the eigenvalues of $\sqrt{\rho\sqrt{\tilde{\rho}}\rho}$, with $\lambda_{1}>\lambda_{2}\geq\lambda_{3}\geq\lambda_{4}$. Here, $\tilde{\rho} = (\sigma_{y}\otimes\sigma_{y})\rho^*(\sigma_{y}\otimes\sigma_{y})$, with the matrix $\sigma_{y}=[[0 \, -i];[i \, 0]]$, and $\rho^*$ the element-wise complex conjugation of $\rho$. Alternatively, these eigenvalues can be obtained as the square roots of the eigenvalues of $\rho\tilde{\rho}$. 

In our case, for $\rho_{I,M}$ we find 
\begin{eqnarray}
\lambda_{1} &=& \sqrt{\frac{x}{2}}+|w|, \\
\lambda_{2} &=& \sqrt{\frac{x}{2}}-|w|, \\
\lambda_{3} &=& 0, \\
\lambda_{4} &=& 0,
\end{eqnarray}
and hence, the concurrence is given by $C_{I-M}=\max\{0,2|w|\}$. In the case of $\rho_{I,U}$, we have 
\begin{eqnarray}
\lambda_{1} &=& \frac{\sqrt{1-2x}}{2}+|y|, \\
\lambda_{2} &=& \frac{\sqrt{1-2x}}{2}-|y|, \\
\lambda_{3} &=& 0, \\
\lambda_{4} &=& 0,
\end{eqnarray}
finding the concurrence $C_{I-U}=\max\{0,2|y|\}$. Therefore, the concurrences can be computed as
\begin{eqnarray}
C_{I,M} &=& \Big|\cos G_{U}|\xi|t \Big|, \\
C_{I,U} &=& \Big|\sin G_{U}|\xi|t \Big|.
\end{eqnarray}

The concurrence of a state with density matrix $\rho$ is used to obtain the well-known entanglement of formation for the same state, $\mathcal{E}_{\rho}$, in the two-qubit mixed-state scenario. This is done by computing~\cite{Wootters2001} 
\begin{equation}
\mathcal{E}_{\rho} = 1 - \frac{1}{2}\left[ \left( 1+\sqrt{1-C^{2}}\right)\log_{2}\left( 1+\sqrt{1-C^{2}}\right) + \left( 1-\sqrt{1-C^{2}}\right)\log_{2}\left( 1-\sqrt{1-C^{2}}\right) \right].
\end{equation}

\section{Concurrence without semiclassical approximation for spins}\label{sm_6}
In this section, we will look into the semiclassical approximation for spin operators that we performed before computing the concurrence between the idler-mixing and the idler-up-converted frequency modes. This approximation is based on the same one for bosonic states, in which the system is assumed to be in a coherent state $|\alpha\rangle$, and thus we replace $a\rightarrow\alpha$. This assumes that the system remains in a coherent state, which is what we are considering here with the molecules; after these are in a coherent state, we are looking at the dynamics after introducing a single-excitation into the system. Let us derive here the idler-mixing and idler-up-converted states without the semiclassical approximation of the spin operators. We start from the state 
\begin{equation}
\frac{1}{\sqrt{2}}\left(1+|s|^{2}\right)^{-\frac{N}{2}}\sum_{n=0}^{N}\begin{pmatrix} N \\ n\end{pmatrix}^{1/2} \left(s e^{-i\Omega t}\right)^{n}(|0,0,0\rangle+|1,1,0\rangle)\left|{\bf \frac{N}{2}},-\frac{N}{2}+n\right\rangle
\end{equation}
in the Schr\"{o}dinger picture, which describes the idler, mixing, up-converted, and molecules, in that order. We want to obtain the evolution of this state under our Hamiltonian
\begin{equation}
H_{\text{eff}} = \hbar\Omega J_{z} + \hbar\omega_{M}a_{M}^{\dagger}a_{M} + \hbar \omega_{U}a_{U}^{\dagger}a_{U} + \hbar G_{U}\left( e^{i\Delta\beta z}a_{M}^{\dagger} a_{U}J^{+}  + e^{-i\Delta\beta z}a_{M} a_{U}^{\dagger}J^{-}\right), 
\end{equation}
where we have already implemented the phase-matching condition $\beta_{U}-\beta_{M}=\Delta\beta$. Again, we now move to an interaction picture with respect to $\hbar\Omega J_{z} + \hbar\omega_{M}a_{M}^{\dagger}a_{M} + \hbar \omega_{U}a_{U}^{\dagger}a_{U}$, where we have
\begin{equation}
H_\text{eff}^{I} = \hbar G_{U}\left( e^{i\Delta\beta z}a_{M}^{\dagger} a_{U}J^{+}  + e^{-i\Delta\beta z}a_{M} a_{U}^{\dagger}J^{-}\right).
\end{equation}
Here, the time-dependence of the Hamiltonian is cancelled due to the resonance $\Omega-\omega_{U}+\omega_{M}=0$, as we have seen before. The initial state in this interaction picture becomes
\begin{equation}
\frac{1}{\sqrt{2}}\left(\frac{e^{-i\Omega t}}{1+|s|^{2}}\right)^{\frac{N}{2}}\sum_{n=0}^{N}\begin{pmatrix} N \\ n\end{pmatrix}^{1/2} s^{n}\left(|0,0,0\rangle+e^{i\omega_{M}t}|1,1,0\rangle\right)\left|{\bf \frac{N}{2}},-\frac{N}{2}+n\right\rangle.
\end{equation}
First of all, we can easily see that $H_\text{eff}^{I}|0,0,0\rangle\left|{\bf \frac{N}{2}},-\frac{N}{2}+n\right\rangle = 0$. Then, we need to compute 
\begin{eqnarray}
&& H_\text{eff}^{I}|1,1,0\rangle\left|{\bf \frac{N}{2}},-\frac{N}{2}+n\right\rangle = \hbar G_{U}e^{-i\Delta\beta z}\sqrt{n(N-n+1)}|1,0,1\rangle\left|{\bf \frac{N}{2}},-\frac{N}{2}+n-1\right\rangle, \\
&& H_\text{eff}^{I}|1,0,1\rangle\left|{\bf \frac{N}{2}},-\frac{N}{2}+n-1\right\rangle = \hbar G_{U}e^{i\Delta\beta z}\sqrt{n(N-n+1)}|1,1,0\rangle\left|{\bf \frac{N}{2}},-\frac{N}{2}+n\right\rangle.
\end{eqnarray}
Again, the single-excitation subspace is closed, and can be diagonalized for a fixed $n$, finding that the energies are $E_{n}^{\pm}=\pm\hbar G_{U}\sqrt{n(N-n+1)}$. In this case, we try the eigenstates $|\phi_{n}^{\pm}\rangle = a|1,1,0\rangle\left|{\bf \frac{N}{2}},-\frac{N}{2}+n\right\rangle \pm b |1,0,1\rangle\left|{\bf \frac{N}{2}},-\frac{N}{2}+n-1\right\rangle$ in the eigenvalue equation, to find
\begin{equation}
|\phi_{n}^{\pm}\rangle = \frac{1}{\sqrt{2}}\left(|1,1,0\rangle\left|{\bf \frac{N}{2}},-\frac{N}{2}+n\right\rangle \pm e^{-i\Delta\beta z} |1,0,1\rangle\left|{\bf \frac{N}{2}},-\frac{N}{2}+n-1\right\rangle \right).
\end{equation}
Now, we can compute the evolution of the initial state under the Hamiltonian $H_\text{eff}^{I}$, obtaining
\begin{eqnarray}
&& \frac{1}{\sqrt{2}}\left(\frac{e^{-i\Omega t}}{1+|s|^{2}}\right)^{\frac{N}{2}}\sum_{n=0}^{N}\begin{pmatrix} N \\ n\end{pmatrix}^{1/2} s^{n}\left[ |0,0,0\rangle\bigg|{\bf \frac{N}{2}},-\frac{N}{2}+n\right\rangle + \\
\nonumber && + e^{i\omega_{M}t}\cos(G_{U}t\sqrt{n(N-n+1)})|1,1,0\rangle\left|{\bf \frac{N}{2}},-\frac{N}{2}+n\right\rangle \\
\nonumber && -i e^{i(\omega_{M}t-\Delta\beta z)}\sin(G_{U}t\sqrt{n(N-n+1)})|1,0,1\rangle\left|{\bf \frac{N}{2}},-\frac{N}{2}+n-1\right\rangle \bigg].
\end{eqnarray}
Going back to the Schr\"{o}dinger picture, this state is expressed as
\begin{eqnarray}
&& \frac{1}{\sqrt{2}}\left(1+|s|^{2}\right)^{-\frac{N}{2}}\sum_{n=0}^{N}\begin{pmatrix} N \\ n\end{pmatrix}^{1/2} \left(se^{-i\Omega t}\right)^{n} \left[ |0,0,0\rangle\bigg|{\bf \frac{N}{2}},-\frac{N}{2}+n\right\rangle + \\
\nonumber && + \cos(G_{U}t\sqrt{n(N-n+1)})|1,1,0\rangle\left|{\bf \frac{N}{2}},-\frac{N}{2}+n\right\rangle -i e^{-i \Delta\beta z}\sin(G_{U}t\sqrt{n(N-n+1)})|1,0,1\rangle\left|{\bf \frac{N}{2}},-\frac{N}{2}+n-1\right\rangle \bigg].
\end{eqnarray}
The density matrix that we obtain after performing a partial trace operation with respect to the molecular subsystem is
\begin{eqnarray}
\rho &=& \sum_{n=0}^{N} \Big[ \big( c_{0}(n)|0,0,0\rangle + c_{1}(n)|1,1,0\rangle\big)\big( c_{0}^*(n)\langle0,0,0| + c_{1}^*(n)\langle1,1,0|\big) + |c_{2}(n)|^{2}|1,0,1\rangle\langle1,0,1| \Big] + \\
\nonumber && \sum_{n=0}^{N-1} \Big[ \big( c_{0}(n)|0,0,0\rangle + c_{1}(n)|1,1,0\rangle\big)c_{2}^*(n+1)\langle1,0,1| + c_{2}(n+1)|1,0,1\rangle\big( c_{0}^*(n)\langle0,0,0| + c_{1}^*(n)\langle1,1,0|\big) \Big],
\end{eqnarray}
where we have defined
\begin{eqnarray}
c_{0}(n) &=& \frac{1}{\sqrt{2}}\left(1+|s|^{2}\right)^{-\frac{N}{2}}\begin{pmatrix} N \\ n\end{pmatrix}^{1/2} \left(se^{-i\Omega t}\right)^{n}, \\
c_{1}(n) &=& \frac{1}{\sqrt{2}}\left(1+|s|^{2}\right)^{-\frac{N}{2}}\begin{pmatrix} N \\ n\end{pmatrix}^{1/2} \left(se^{-i\Omega t}\right)^{n}\cos(G_{U}t\sqrt{n(N-n+1)}), \\
c_{2}(n) &=& \frac{1}{\sqrt{2}}\left(1+|s|^{2}\right)^{-\frac{N}{2}}\begin{pmatrix} N \\ n\end{pmatrix}^{1/2} \left(se^{-i\Omega t}\right)^{n} (-i) e^{-i\Delta\beta z} \sin(G_{U}t\sqrt{n(N-n+1)}).
\end{eqnarray}
By tracing the up-converted subsystem, we obtain the idler-mixing density matrix,
\begin{equation}
\rho_{I,M} = \sum_{n=0}^{N} \Big[ \big( c_{0}(n)|0,0\rangle + c_{1}(n)|1,1\rangle\big)\big( c_{0}^*(n)\langle0,0| + c_{1}^*(n)\langle1,1|\big) + |c_{2}(n)|^{2}|1,0\rangle\langle1,0| \Big],
\end{equation}
while the density matrix for the idler and the up-converted frequency modes is obtained by tracing the mixing subsystem,
\begin{eqnarray}
\rho_{I,U} &=& \sum_{n=0}^{N} \Big[ |c_{0}(n)|^{2} |0,0\rangle\langle0,0| + |c_{1}(n)|^{2} |1,0\rangle\langle1,0| + |c_{2}(n)|^{2}|1,1\rangle\langle1,1| \Big] + \\
\nonumber && \sum_{n=0}^{N-1} \Big[ \big( c_{0}(n) c_{2}^*(n+1)|0,0\rangle\langle1,1| + c_{2}(n+1) c_{0}^*(n)|1,1\rangle\langle0,0| \Big].
\end{eqnarray}
Expressing these in matrix form, we obtain
\begin{eqnarray}
\rho_{I,M} &=& \begin{pmatrix} x & 0 & 0 & w \\ 0 & \frac{1}{2} - x & 0 & 0 \\ 0 & 0 & 0 & 0 \\ w^* & 0 & 0 & \frac{1}{2} \end{pmatrix}, \\[1ex]
\rho_{I,U} &=& \begin{pmatrix} \frac{1}{2} - x & 0 & 0 & y \\ 0 & x & 0 & 0 \\ 0 & 0 & 0 & 0 \\ y^* & 0 & 0 & \frac{1}{2} \end{pmatrix},
\end{eqnarray}
having identified
\begin{eqnarray}
x &=& \sum_{n=0}^{N} |c_{1}(n)|^{2} = \frac{1}{2}(1+|s|^{2})^{-N}\sum_{n=0}^{N}\begin{pmatrix} N \\ n\end{pmatrix}|s|^{2n}\cos^{2}(G_{U}t\sqrt{n(N-n+1)}), \\
w &=& \sum_{n=0}^{N} c_{1}(n) c_{0}^*(n) = \frac{1}{2}(1+|s|^{2})^{-N} \sum_{n=0}^{N}\begin{pmatrix} N \\ n\end{pmatrix}|s|^{2n}\cos(G_{U}t\sqrt{n(N-n+1)}), \\
y &=& \sum_{n=0}^{N-1} c_{2}(n+1) c_{0}^*(n) = -i e^{-i(\Omega t + \Delta\beta z)}\frac{s}{2}(1+|s|^{2})^{-N}\sum_{n=0}^{N-1}\begin{pmatrix} N \\ n\end{pmatrix}\sqrt{\frac{N-n}{n+1}} |s|^{2n}\sin(G_{U}t\sqrt{(n+1)(N-n)}).
\end{eqnarray}
Now we would like to compare the output states with and without the semiclassical approximation for the molecules. To do that, we could look at, for example, the value of $w$ in both cases:
\begin{eqnarray}
&& w = \frac{1}{2}(1+|s|^{2})^{-N}\sum_{n=0}^{N}\begin{pmatrix} N \\ n\end{pmatrix}|s|^{2n}\cos(G_{U}t\sqrt{n(N-n+1)}), \\
&& w_{\text{sc}} = \frac{1}{2}\cos G_{U}|\xi|t.
\end{eqnarray}
We now take the cosine and expand it in its Taylor series, to write
\begin{equation}
w = \frac{1}{2}(1+|s|^{2})^{-N}\sum_{m=0}^{\infty}(-1)^{m}\frac{(G_{U}t)^{2m}}{(2m)!}\sum_{n=0}^{N}\begin{pmatrix} N \\ n\end{pmatrix}|s|^{2n}n^{m}(N-n+1)^{m}.
\end{equation}
If we try to solve the sum over $n$, we will see that the leading power of $N$ goes as
\begin{equation}\label{sum_approx}
(1+|s|^{2})^{-N}\sum_{n=0}^{N}\begin{pmatrix} N \\ n\end{pmatrix}|s|^{2n}n^{m}(N-n+1)^{m} \sim \left( \frac{N |s|}{1+|s|^{2}}\right)^{2m}.
\end{equation}
Therefore, plugging this back into our previous equation, we have
\begin{equation}
w = \frac{1}{2}\sum_{m=0}^{\infty}(-1)^{m}\frac{(G_{U}t)^{2m}}{(2m)!}\left( \frac{N |s|}{1+|s|^{2}}\right)^{2m} = \frac{1}{2}\cos\left( G_{U} t  \frac{N |s|}{1+|s|^{2}}\right).
\end{equation}
Furthermore, notice that 
\begin{equation}
\frac{N |s|}{1+|s|^{2}} = N \frac{|\tan(G_{S}\alpha_{P}\alpha_{S} t)|}{1+\tan^{2}(G_{S}\alpha_{P}\alpha_{S} t)} = \frac{N}{2}\left| \sin\left(2G_{S}\alpha_{P}\alpha_{S} t\right)\right| = |\xi|.
\end{equation} 
This way, we recover the value of $w$ in the semiclassical approximation of the molecules. Thus, this approximation is valid in the case in which $N$ is large, where we can approximate the sum in Eq.~\eqref{sum_approx}. If we compute the next order in the series, that is, the term that goes with $N^{2m-1}$, and solve the sum, we find 
\begin{equation}
\frac{G_{U}t}{8}\Big[ -N G_{U}t\cos^{2}(2G_{S}\alpha_{P}\alpha_{S} t)\cos G_{U}|\xi|t + \left[\cot(2G_{S}\alpha_{P}\alpha_{S} t) - \tan(2G_{S}\alpha_{P}\alpha_{S} t)\right]\sin G_{U}|\xi|t \Big]. 
\end{equation}
This is very small, since the leading order is $G_{U}t \sim 10^{-17}$. We can also see this as
\begin{equation}
\frac{1}{8N}\left[ -G_{U}^{2}t^{2}|\xi|^{2}\left(1-\cot^{2}(2G_{S}\alpha_{P}\alpha_{S} t)\right)^{2}\cos G_{U}|\xi|t + G_{U}t |\xi |\left(\frac{1}{\sin^{2}(2G_{S}\alpha_{P}\alpha_{S} t)} - \frac{3}{\cos^{2}(2G_{S}\alpha_{P}\alpha_{S} t)}\right)\sin G_{U}|\xi|t \right]. 
\end{equation}
Given that $G_{U}t|\xi| \sim 1$, we have that this term is led by a factor of $1/N$, and therefore should be small. In the limit $t\rightarrow0$, this term goes to zero.
\begin{figure}[t]
    \centering
    \includegraphics[width=0.5\linewidth]{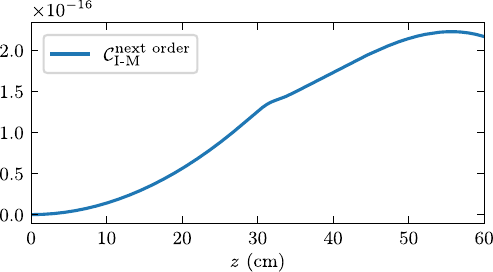}
    \caption{The next order in the idler-mixing concurrence summation plotted as a function of the propagation distance $z$ (see Eq.~\eqref{C_IM_next_order}). The parameters for this simulation are the same ones used for the figures in the main text.}
    \label{fig_s3}
\end{figure}
The next-order correction to the idler-mixing concurrence is
\begin{equation}\label{C_IM_next_order}
C_{I,M}^{\text{next order}} = \left| \frac{G_{U}t}{4}\Big[ -N G_{U}t\cos^{2}(2G_{S}\alpha_{P}\alpha_{S} t)\cos G_{U}|\xi|t + \left[\cot(2G_{S}\alpha_{P}\alpha_{S} t) - \tan(2G_{S}\alpha_{P}\alpha_{S} t)\right]\sin G_{U}|\xi|t \Big]\right|.
\end{equation}
This is represented in Fig.~\ref{fig_s3} against the length of the fiber, where we can observe that these corrections are very small, reaching maximum values of $\sim2\times10^{-16}$ during the length of the fiber.

\section{Equations of motion}\label{sm_7}
Going back to the single-molecule case, here is shown that the Maxwell-Bloch equations describing stimulated Raman scattering~\cite{SM_Raymer1990} can be recovered through the equations of motion associated to the effective Hamiltonian derived in the previous section. In the following, we only describe the pump-Stokes interaction, considering that the anti-Stokes population is negligible throughout the process. We will work with the Hamiltonian
\begin{eqnarray}\label{H_Stokes_eff}
H_{\text{eff}} &=& \frac{\hbar\Omega}{2}\sigma_{z} + \hbar\Big[ (\omega_{P}+\Delta_{P}^{+})\mathbb{1}_{2} + \Delta_{P}^{-}\sigma_{z}\Big]a^{\dagger}_{P}a_{P} + \hbar\Big[ (\omega_{S}+\Delta_{S}^{+})\mathbb{1}_{2} + \Delta_{S}^{-}\sigma_{z}\Big]a^{\dagger}_{S}a_{S} \\
\nonumber &+& \hbar \left(G_{S} e^{i\Delta\beta z}a_{P}a^{\dagger}_{S}\sigma^{+} + G_{S}^{*}e^{-i\Delta\beta z}a^{\dagger}_{P}a_{S}\sigma^{-}\right).
\end{eqnarray}
Here we will derive the equations of motion (EOM) that describe the change in populations in a single molecule due to the interaction with the pump and Stokes fields. First, we will start from the Stokes effective Hamiltonian, apply the semiclassical approximation on the pump and Stokes fields, and obtain the EOM. Note that these will include both a semiclassical and an adiabatic approximation. Another approach follows from the original Hamiltonian, applying a semiclassical approximation, and obtain the EOM, to finally apply the adiabatic approximation. The latter was the path taken in previous works, while the former is the path we want to follow. Our goal is to show that they are both equivalent.

We first start in the Schr\"{o}dinger picture, and go to an interaction picture with respect to the pump and Stokes fields. We transform $H_{\text{eff}}$ by $\omega_{P}a^{\dagger}_{P}a_{P} + \omega_{S}a^{\dagger}_{S}a_{S}$,
obtaining the interaction picture Hamiltonian
\begin{eqnarray}\label{H_Stokes_eff}
H_{\text{eff}}^{I}(t) &=&  \frac{\hbar\Omega}{2}\sigma_{z} + \hbar\left(\Delta^{+}_{P}\mathbb{1}_{2} + \Delta_{P}^{-}\sigma_{z}\right) a^{\dagger}_{P}a_{P}+ \hbar\left(\Delta^{+}_{S}\mathbb{1}_{2} + \Delta_{S}^{-}\sigma_{z}\right) a^{\dagger}_{S}a_{S} \\
\nonumber &+& \hbar\left(G_{S} e^{i\left(\Delta\beta z - \Omega t\right)}a_{P}a^{\dagger}_{S}\sigma^{+}+ G_{S}^{*}e^{-i\left(\Delta\beta z - \Omega t\right)}a^{\dagger}_{P}a_{S}\sigma^{-}\right).
\end{eqnarray}
Now, we apply a semiclassical approximation in the bosonic modes, which amounts to replacing the associated creation and annihilation operators by their expectation values over a coherent state. That is, $a_{l}\rightarrow\alpha_{l}$ and $a_{l}^{\dagger}\rightarrow\alpha_{l}^{*}$ for $l\in\{P,S\}$,
\begin{eqnarray}
H_{\alpha}^{I}(t) &=&  \frac{\hbar\Omega}{2}\sigma_{z} + \hbar\left(\Delta^{+}_{P}\mathbb{1}_{2} + \Delta_{P}^{-}\sigma_{z}\right) |\alpha_{P}|^{2}+ \hbar\left(\Delta^{+}_{S}\mathbb{1}_{2} + \Delta_{S}^{-}\sigma_{z}\right) |\alpha_{S}|^{2} \\
\nonumber &+& \hbar\left(G_{S} e^{i\left(\Delta\beta z - \Omega t\right)}\alpha_{P}\alpha_{S}^{*}\sigma^{+}+ G_{S}^{*}e^{-i\left(\Delta\beta z - \Omega t\right)}\alpha_{P}^{*}\alpha_{S}\sigma^{-}\right).
\end{eqnarray}
The equations of motion for the state of the molecule, $\rho$, are computed using the von Neumann equation, $\hbar\dot{\rho}=i[\rho,H]$. We can split $\rho$ into its basis operators,
\begin{equation}
\rho = \rho_{0,0}\left(\frac{\mathbb{1}_{2}-\sigma_{z}}{2}\right) +\rho_{0,1}\sigma^{-} +\rho_{1,0}\sigma^{+} + \rho_{1,1}\left(\frac{\mathbb{1}_{2}+\sigma_{z}}{2}\right) = \mathbb{1}_{2} + (\rho_{1,1}-\rho_{0,0})\frac{\sigma_{z}}{2}+\rho_{0,1}\sigma^{-} +\rho_{1,0}\sigma^{+},
\end{equation}
since $\rho_{1,1}+\rho_{0,0}=\tr\rho=1$, and express $[\rho,H]$ in such basis. For that, we compute
\begin{eqnarray}
\nonumber [\sigma_{z},H] &=& 2\hbar\left(G_{S} e^{i\left(\Delta\beta z - \Omega t\right)}\alpha_{P}\alpha_{S}^{*}\sigma^{+}-G_{S}^{*} e^{-i\left(\Delta\beta z - \Omega t\right)}\alpha_{P}^{*}\alpha_{S}\sigma^{-}\right), \\
\nonumber [\sigma^{+},H] &=& -\hbar\left(\Omega + 2\Delta_{P}^{-}|\alpha_{P}|^{2} + 2\Delta_{S}^{-}|\alpha_{S}|^{2}\right)\sigma^{+} + \hbar G_{S}^{*}e^{-i\left(\Delta\beta z - \Omega t\right)}\alpha_{P}^{*}\alpha_{S}\sigma_{z}, \\
\nonumber [\sigma^{-},H] &=& \hbar\left(\Omega + 2\Delta_{P}^{-}|\alpha_{P}|^{2} + 2\Delta_{S}^{-}|\alpha_{S}|^{2}\right)\sigma^{-} - \hbar G_{S} e^{i\left(\Delta\beta z - \Omega t\right)}\alpha_{P}\alpha_{S}^{*}\sigma_{z},
\end{eqnarray}
using the formulas
\begin{eqnarray}
\nonumber [\sigma_{z},\sigma^{+}] &=& 2\sigma^{+}, \\
\nonumber [\sigma_{z},\sigma^{-}] &=& -2\sigma^{-}, \\
\nonumber [\sigma^{+},\sigma^{-}] &=& \sigma_{z}.
\end{eqnarray}
By defining $w=\rho_{1,1}-\rho_{0,0}$, we write the following EOM,
\begin{eqnarray}
\dot{w} &=& -2i\left(\rho_{0,1}(t)G_{S} e^{i\left(\Delta\beta z - \Omega t\right)} \alpha_{P}\alpha_{S}^{*} - \rho_{1,0}(t)G_{S}^{*}e^{-i\left(\Delta\beta z - \Omega t\right)}\alpha_{P}^{*}\alpha_{S}\right), \\
\dot{\rho}_{0,1} &=& i\rho_{0,1}(t)\left(\Omega + 2\Delta_{P}^{-}|\alpha_{P}|^{2} + 2\Delta_{S}^{-}|\alpha_{S}|^{2}\right) - iw(t)G_{S}^{*}e^{-i\left(\Delta\beta z - \Omega t\right)}\alpha_{P}^{*}\alpha_{S}, \\
\dot{\rho}_{1,0} &=& -i\rho_{1,0}(t)\left(\Omega + 2\Delta_{P}^{-}|\alpha_{P}|^{2} + 2\Delta_{S}^{-}|\alpha_{S}|^{2}\right) + iw(t)G_{S} e^{i\left(\Delta\beta z - \Omega t\right)}\alpha_{P}\alpha_{S}^{*}.
\end{eqnarray}
Generally, the $\Delta_{l}^{-}$ terms are Stark shifts that can be neglected~\cite{SM_Raymer1990}. In order for this result to match the Maxwell-Bloch equations~\cite{SM_Raymer1990,Bauerschmidt2016}, we need to identify $G_{S}\alpha_{P}\alpha_{S}^{*} = -\kappa_{1,p}E_{P}E_{S}^{*}$. Using this relation, we estimate $G_{S}$ in terms of the phenomenological factor $\kappa_{1,p}$. This leaves us with
\begin{eqnarray}
\dot{w} &=& 2i\left(\rho_{0,1}(t)\kappa_{1,p} E_{P}E_{S}^{*}e^{i\left(\Delta\beta z - \Omega t\right)} - \rho_{1,0}(t)\kappa^{*}_{1,p}E_{P}^{*}E_{S}e^{-i\left(\Delta\beta z - \Omega t\right)}\right), \\
\dot{\rho}_{0,1} &=& i\rho_{0,1}(t)\Omega + i w(t) \kappa_{1,p}^{*}E_{P}^{*}E_{S}e^{-i\left(\Delta\beta z - \Omega t\right)},
\end{eqnarray}
where we have not written the equation for $\dot{\rho}_{10}$ because it is just the complex conjugate of that for $\dot{\rho}_{0,1}$. The differences we find with the previous results lie in the definition of the electric field. While we have defined $\mathcal{E}(z,t) = \mathcal{E}_{P}(z,t) + \mathcal{E}_{S}(z,t)$, previous works defined $\mathcal{E}(z,t) = (\mathcal{E}_{P}(z,t) + \mathcal{E}_{S}(z,t))/2$, and thus the extra factor of $1/4$. Furthermore, while we have chosen the plane-wave solution of the electric fields as
\begin{equation}
\mathcal{E}_{l}(z,t) = E_{l}e^{i(\beta_{l}z - \omega_{l}t)} + E_{l}^{*}e^{-i(\beta_{l}z - \omega_{l}t)},
\end{equation}
previous works have chosen
\begin{equation}
\mathcal{E}_{l}(z,t) = E_{l}e^{-i(\beta_{l}z - \omega_{l}t)} + E_{l}^{*}e^{i(\beta_{l}z - \omega_{l}t)},
\end{equation}
and this is why we find $E_{P}^{*}E_{S}$ instead of $E_{P}E_{S}^{*}$.

\end{document}